\documentclass[twocolumn]{aastex701}

\begin{document}

\title{The Nearby Star Formation and Supernova Histories Reconstructed from Young Star Clusters}

\author[orcid=0000-0001-9201-5995,sname=Swiggum,gname=Cameren]{Cameren Swiggum}
\affiliation{Center for Astrophysics | Harvard \& Smithsonian, 60 Garden Street, Cambridge, MA 02138, USA}
\email{cameren.swiggum@cfa.harvard.edu}
\correspondingauthor{Cameren Swiggum}

\author[orcid=0000-0002-2250-730X]{Catherine Zucker}
\affiliation{Center for Astrophysics | Harvard \& Smithsonian, 60 Garden Street, Cambridge, MA 02138, USA}
\email{catherine.zucker@cfa.harvard.edu}

\author[orcid=0000-0001-9933-1229]{Michelangelo Pantaleoni González}
\affiliation{Department of Astrophysics, University of Vienna, Türkenschanzstraße 17, 1180 Vienna, Austria}
\email{michelangelo.pantaleoni@univie.ac.at}

\author[orcid=0000-0002-5555-8058]{Emily L. Hunt}
\affiliation{Department of Astrophysics, University of Vienna, Türkenschanzstraße 17, 1180 Vienna, Austria}
\email{emily.lauren.hunt@univie.ac.at}

\author[orcid=0000-0002-8109-2642]{Robert A. Benjamin}
\affiliation{Department of Physics, University of Wisconsin--Whitewater, Whitewater, WI 53190, USA}
\email{benjamir@uww.edu}

\author[orcid=0000-0002-6952-9688]{Sebastian Hutschenreuter}
\affiliation{Department of Astrophysics, University of Vienna, Türkenschanzstraße 17, 1180 Vienna, Austria}
\email{sebastian.hutschenreuter@univie.ac.at}

\author[orcid=0000-0002-9817-7095]{Alena K. Rottensteiner}
\affiliation{Department of Astrophysics, University of Vienna, Türkenschanzstraße 17, 1180 Vienna, Austria}
\email{alena.kristina.rottensteiner@univie.ac.at}

\author{Efrem Maconi}
\affiliation{Department of Astrophysics, University of Vienna, Türkenschanzstraße 17, 1180 Vienna, Austria}
\email{efrem.maconi@univie.ac.at}

\author[orcid=0009-0007-1181-8034]{Lewis McCallum}
\affiliation{Center for Astrophysics | Harvard \& Smithsonian, 60 Garden Street, Cambridge, MA 02138, USA}
\email{lewis.mccallum@cfa.harvard.edu}

\author[orcid=0000-0002-4355-0921]{João Alves}
\affiliation{Department of Astrophysics, University of Vienna, Türkenschanzstraße 17, 1180 Vienna, Austria}
\email{joao.alves@univie.ac.at}

\author[orcid=0000-0002-6215-4270]{Sebastian Ratzenböck}
\affiliation{Center for Astrophysics | Harvard \& Smithsonian, 60 Garden Street, Cambridge, MA 02138, USA}
\email{sebastian.ratzenboeck@cfa.harvard.edu}

\begin{abstract}
We reconstruct the recent star formation and core-collapse supernova (ccSN) histories of the Solar Neighborhood from the past trajectories of young star clusters. Using a \textit{Gaia}-based cluster sample with newly derived ages, masses, and bulk 3D velocities, we integrate orbits backward in an assumed axisymmetric Galactic potential and combine the trajectories with IMF sampling and stellar lifetimes to infer ccSN times and locations over the past 50 Myr. The result is an all-sky, 3D, time-resolved map of nearby ccSN activity for comparison with high-resolution 3D views of the local interstellar medium. The 0--15 Myr map shows strong enhancements toward Orion, Vela, Sco--Cen, and Cepheus, many within present-day cavities and shells. At earlier times, the dominant enhancements trace the Collinder 135, Messier 6, and Alpha Persei cluster families, showing how the remnants of massive star-forming complexes have shaped the recent local feedback history. We recover a bursty star formation history followed by a delayed, smoother ccSN history. Over the last 40 Myr, the mean star formation and ccSN rates are \(823~M_\odot~\mathrm{Myr}^{-1}\) and \(7.7~\mathrm{Myr}^{-1}\), respectively, corresponding to a Milky Way rate of \(0.55\pm0.03~\mathrm{century}^{-1}\). Present-day OB-star catalogs yield rates ranging from agreement with the cluster reconstruction to several times higher. Because the catalogs overlap weakly and require different corrections, we do not rescale the ccSN map. Our reconstruction provides an empirical framework for connecting the recent history of massive-star feedback to the 3D structure and life cycle of gas in the nearby Milky Way.
\end{abstract}

\keywords{open star clusters and associations --- supernovae: general --- ISM: bubbles --- ISM: structure --- solar neighborhood --- Galaxy: kinematics and dynamics}

\section{Introduction}\label{sec:intro}

Core-collapse supernovae (ccSNe) are a dominant mechanical energy source for the interstellar medium (ISM). Each event releases $\sim 10^{51}$~erg, and clustered explosions can inflate superbubbles, drive turbulence and multiphase structure, and launch outflows that regulate the cycling of gas through galactic disks \citep{kim_momentum_2015,walch_silcc_2015,kim_superbubbles_2017,fielding_clustered_2018}. Multi-wavelength surveys of nearby galaxies, aided by JWST's infrared sensitivity, resolve feedback-driven bubbles, shells, and cavities across a range of galactic environments \citep{watkins_phangsjwst_2023,barnes_phangsjwst_2023}.

The Solar Neighborhood is the only laboratory where the ISM and the stellar populations responsible for shaping it can both be resolved in three dimensions (3D). With precise astrometry for more than one billion stars, \textit{Gaia} has made it possible to map the local ISM in 3D. These 3D dust maps now trace nearby molecular clouds, cavities, shell surfaces, and kiloparsec-scale star-forming structures with parsec-scale resolution \citep{lallement_gaia_2019,vergely_three-dimensional_2022,edenhofer_parsec-scale_2024,alves_galactic-scale_2020,swiggum_radcliffe_2022,konietzka_radcliffe_2024,kormann_superclouds_2025,oneill_local_2024}.

\textit{Gaia}-based star cluster catalogs also provide ages, masses, and 3D motions for nearby clusters \citep{cantat-gaudin_gaia_2018,castro-ginard_hunting_2022,kerr_spyglass_2023,kerr_spyglass_2026,hunt_improving_2023,hunt_improving_2024}. The orbits of young star clusters can be integrated backwards in the Galaxy to map their birth locations. Using this approach, \citet{swiggum_most_2024} showed that most young clusters within $\sim1~$kpc trace back to three past, massive complexes that formed stars over the last 60 Myr. These cluster families now lie within large present-day dust cavities, linking their feedback history to the structure of the local ISM \citep{vergely_three-dimensional_2022}.

One of these families, Alpha Persei, includes the young Sco--Cen population whose supernovae likely helped form the Local Bubble \citep{swiggum_most_2024}. This agrees with earlier \textit{Hipparcos}-era tracebacks that linked Sco--Cen supernova activity to the Bubble's origin \citep{maiz-apellaniz_origin_2001,berghofer_origin_2002,fuchs_search_2006}. Nearby supernovae have also been invoked to explain terrestrial and lunar \(^{60}\mathrm{Fe}\) signatures \citep{wallner_recent_2016,breitschwerdt_locations_2016,siegert_gamma-ray_2024}.

In this work, we extend the cluster traceback approach to a sample of 568 young star clusters, defined by present-day positions satisfying $|X|,|Y|,|Z|\leq1.25$ kpc and chosen for comparison with the 3D dust map of \citet{edenhofer_parsec-scale_2024}. Building on \citet{swiggum_most_2024}, we use the past trajectories of these star clusters, together with their ages and masses, to infer the expected ccSN times and locations across the Solar Neighborhood using stellar evolution models and IMF sampling. The result is an all-sky, 3D, time-resolved reconstruction of nearby ccSN activity over the past 50 Myr that can be compared with the newly available, high-resolution 3D view of the local ISM and inform data-driven simulations of the Local Bubble and surrounding ISM \citep{romano_sissi_2025-1,romano_sissi_2025}.

\S\ref{sec:data} defines the cluster samples, radial velocity compilation, and OB star catalogs used for comparison. \S\ref{sec:methods} describes the age and mass fitting, orbit tracebacks, IMF sampling, ccSN-position assignment, and validation against present-day massive star counts. \S\ref{sec:results} presents the recent and time-resolved ccSN reconstructions, the inferred star formation and ccSN rate histories, the time dependence of ccSN clustering, and the OB star comparison. \S\ref{sec:discussion} discusses connections between reconstructed ccSNe and nearby ISM shells, regional feedback histories, the clustering and burstiness of recent feedback, uncertainties in the absolute star formation and ccSN rates, and the main limitations of the reconstruction.

\section{Data}\label{sec:data}

\subsection{Star cluster catalog}\label{sec:data_catalog}

We use the all-sky open cluster catalog of \citet{hunt_improving_2023} (hereafter HR23), who identified 7167 star clusters in \textit{Gaia} DR3 using the Hierarchical Density-Based Spatial Clustering of Applications with Noise algorithm, HDBSCAN \citep{hutchison_density-based_2013,mcinnes_hdbscan_2017}. The catalog includes both previously known objects and newly identified candidates. Candidate clusters were further characterized with astrometric parameters, distances, extinctions, ages, membership lists, a cluster significance test score, and a color--magnitude-diagram classification probability.

We construct our cluster sample from the high-reliability subset recommended by \citet{hunt_improving_2023}. Specifically, we require $\texttt{class\_50} > 0.5$, corresponding to a median color--magnitude-diagram classification probability consistent with a single stellar population, and $\texttt{cst} > 5$, corresponding to a $>5\sigma$ astrometric overdensity relative to the local field. We then impose a spatial cut of $|X|,|Y|,|Z|\leq1.25$ kpc in heliocentric Galactic Cartesian coordinates, with $X$ directed toward the Galactic center, $Y$ in the direction of Galactic rotation, and $Z$ toward the north Galactic pole. This cut defines the 2.5 kpc cube sample used for the full ccSN map and the comparison to the present-day local ISM. This selection approximately matches the dust-map volume of \citet{edenhofer_parsec-scale_2024} while extending slightly beyond it, where sufficient cluster data remain available.

We use HR23 catalog ages $<200$ Myr only as an initial filtering step. We refit the selected clusters in \S\ref{sec:ages} with \texttt{Chronos}\footnote{\texttt{Chronos} is the informal name used here for the Bayesian isochrone-fitting code. The code is available at \url{https://github.com/sebastianratzenboeck/Chronos}.}, the Bayesian isochrone-fitting method used by \citet{ratzenbock_star_2023}, to obtain the full age posterior for each cluster. These posteriors allow us to identify multimodal solutions and measure the uncertainty associated with the adopted age component. The ages ultimately adopted in the supernova reconstruction are the re-derived \texttt{Chronos} ages, and we retain clusters with \texttt{Chronos} ages $<100$ Myr. Table~\ref{tab:sample} summarizes the main reductions from the parent catalog to the 2.5 kpc cube sample used in the reconstruction.

From these clusters, we define two samples for different parts of the analysis. The 2.5 kpc cube sample is used for the full ccSN map and its comparison with the 3D dust distribution. The 1 kpc cube sample uses present-day cluster positions satisfying $|X|,|Y|,|Z|\leq500$ pc and is used for the rate and clustering statistics, where the Gaia DR3 open cluster census is best characterized. Both samples are restricted by the bulk 3D velocity and re-derived age and mass cuts described in \S\ref{sec:methods}. The 1 kpc cube sample still includes the major recent star-forming environments Orion, Vela, and Sco--Cen \citep{hunt_improving_2024,hunt_selection_2026}.

\subsection{Radial velocities}\label{sec:data_rv}
A cluster's three-dimensional motion requires radial velocities for member stars. We use radial velocities from \textit{Gaia} Data Release 3 \citep{katz_gaia_2023}, the Apache Point Observatory Galactic Evolution Experiment (APOGEE) Data Release 17 \citep{abdurrouf_seventeenth_2022}, the Galactic Archaeology with HERMES (GALAH) Data Release 4 \citep{buder_galah_2025}, the Dark Energy Spectroscopic Instrument (DESI) Milky Way Survey \citep{koposov_desi_2024}, the Gaia--ESO Survey (GES) Data Release 6 \citep{hourihane_gaia-eso_2023}, and the Radial Velocity Experiment (RAVE) Data Release 6 \citep{steinmetz_rave_2020}.

The external spectroscopic catalogs are matched to HR23 member stars using Gaia source identifiers. When a survey contains multiple radial velocity rows for the same Gaia source, we retain the measurement with the smallest reported uncertainty. For stars with valid radial velocities from more than one survey, we adopt an inverse-variance-weighted median of the available survey values. The resulting member star radial velocities are combined with \textit{Gaia} astrometry to estimate one bulk $U,V,W$ velocity for each cluster, as described in \S\ref{sec:bulk_velocities}.

\begin{table*}[t]
\centering
\small
\caption{Cluster sample definition and filtering.\label{tab:sample}}
\resizebox{\textwidth}{!}{%
\begin{tabular}{llr}
\hline
\hline
Stage & Criterion & Retained clusters \\
\hline
Parent catalog & HR23 all-sky open cluster catalog & 7167 \\
2.5 kpc cube high-reliability sample & $\texttt{class\_50}>0.5$, $\texttt{cst}>5$, and $|X|,|Y|,|Z|\leq1.25$ kpc & 1566 \\
Chronos input sample & HR23 catalog age $<200$ Myr & 1166 \\
Bulk velocity sample & $\geq5$ member star radial velocities and $\sigma_U,\sigma_V,\sigma_W<5~{\rm km~s^{-1}}$ & 1011 \\
\hline
2.5 kpc cube sample & Bulk velocity sample with a successful Chronos age/mass fit and adopted Chronos age $<100$ Myr & 568 \\
1 kpc cube sample & 2.5 kpc cube sample with $|X|, |Y|, |Z|\leq500$ pc & 312 \\
\hline
\end{tabular}%
}
\end{table*}

\subsection{Present-day OB star catalogs}\label{sec:data_ob_catalogs}

We compare our cluster-based reconstruction to present-day OB star catalogs to estimate how many massive stars, and therefore how much recent and future ccSN activity, may be missing from our cluster sample. Massive OB stars ($M\geq8\,M_{\odot}$) have lifetimes of about 40 Myr or less and can be cataloged whether they remain in clusters or have dispersed into the field.

\citet{quintana_census_2025} (hereafter Q25) provide a \textit{Gaia}-based census of OB stars within 1 kpc of the Sun. They infer stellar parameters by fitting evolutionary and atmospheric models to multi-band spectral energy distributions, and report masses, distances, and effective temperatures for 24{,}706 stars with $T_{\rm eff} > 10{,}000$ K.

The second catalog, from \citet{pantaleoni_gonzalez_alma_2025}, is the third installment of the Alma Luminous Star catalog (ALS III), following the original all-sky Galactic OB-star compilation of \citet{reed_catalog_2003} and its \textit{Gaia} DR2 update by \citet{pantaleoni_gonzalez_alma_2021}. ALS III contains $15{,}542$ Galactic OB-star candidates with updated \textit{Gaia} DR3 astrometry and photometry, revised distances, and spectroscopic classifications from massive-star surveys.

To place the two observed catalogs and the cluster reconstruction on comparable footing, we select sources intended to trace the same local population of massive OB stars: stars with $M \geq 8\,M_{\odot}$, whose lifetimes are no longer than about 40 Myr. For Q25, we retain stars with inferred masses $M \geq 8\,M_{\odot}$. For ALS III, we select stars flagged as $\texttt{Cat}=\texttt{M}$, which identifies the catalog's $\geq 8\,M_{\odot}$ candidate subset. For the quantitative comparison, we restrict both OB star samples to the 1 kpc cube, $|X|, |Y|, |Z| \leq 500$ pc. Because Q25 is limited to heliocentric distances $d\leq1$ kpc, we restrict the projected spatial comparison of all three populations to this distance and require $|Z|<500$ pc. We use these observed OB star samples to compare against the synthetic massive stars drawn from IMF sampling of the 1 kpc cube sample that are predicted to explode as ccSNe in the future.

\section{Methods}\label{sec:methods}
\subsection{Bulk cluster 3D velocities}\label{sec:bulk_velocities}

We fit each cluster with a single bulk velocity vector by solving directly for the 3D motion that best matches the member star Gaia astrometry and available radial velocities. Most member stars do not have radial velocities, so all stars with usable Gaia positions, parallaxes, and proper motions contribute tangential velocity constraints, while the subset with radial velocities also constrains the line-of-sight motion. We convert each member's proper motion and parallax into two tangential velocity constraints and combine these with the available radial velocities in an iterative, uncertainty-weighted least-squares fit with $5\sigma$ clipping. We solve the resulting linear system for the three Cartesian velocity components with the \texttt{numpy.linalg.lstsq} routine, minimizing the uncertainty-weighted residuals between the observed velocities and the corresponding projections of the common velocity vector. We keep only high-quality velocity fits, defined as fits with at least five radial velocity constraints and component uncertainties below $5~{\rm km~s^{-1}}$. These requirements leave 1011 clusters in the bulk velocity sample (Table~\ref{tab:sample}). For the 568 clusters in the 2.5 kpc cube sample, the mean uncertainties are $0.85$, $0.92$, and $0.54~{\rm km~s^{-1}}$ in $U$, $V$, and $W$, respectively. The corresponding mean uncertainties for the 312 clusters in the 1 kpc cube sample are $0.63$, $0.66$, and $0.53~{\rm km~s^{-1}}$.

\subsection{Cluster Ages and Masses}\label{sec:ages}

HR23 report ages for all clusters in their catalog from a convolutional-neural-network analysis of Gaia color--magnitude diagrams. We use those catalog ages only to define the initial young sample with HR23 ages $<200$ Myr. We then re-fit the selected clusters with \texttt{Chronos}, the Bayesian isochrone-fitting method used by \citet{ratzenbock_star_2023} to recover cluster ages from Gaia photometry. \texttt{Chronos} fits each cluster's Gaia color--magnitude diagram using a single coeval PARSEC isochrone. For this analysis, we fix metallicity to solar and sample the posterior distributions of age, extinction $A_V$, \texttt{skewness}, and \texttt{scale}. The likelihood measures how closely the member stars follow the isochrone. The \texttt{skewness} parameter accounts for preferential offsets toward redder colors and brighter magnitudes, including those caused by unresolved binaries. The \texttt{scale} parameter sets the width of the residual distribution and allows for cluster-to-cluster broadening from effects such as differential reddening and young stellar objects.

We use a flat prior in linear age from 1 Myr to 12 Gyr. This deliberately broad range allows the fit to recover clusters that may be older than inferred by HR23, without restricting their solutions to the initial $<200$ Myr selection. We adopt priors of 0.5--0.99 for \texttt{skewness} and 0.001--0.1 for \texttt{scale}. The $A_V$ prior is informed by three 3D dust maps. We first query E24 \citep{edenhofer_parsec-scale_2024} and multiply its unitless extinction measure, defined by \citet{zhang_parameters_2023}, by 2.8 to obtain $A_V$. Where E24 is unavailable, we query Bayestar19 and multiply its reddening estimates by 2.742 to obtain $A_V$ \citep{green_3d_2019}. Where neither map is available, we query DECaPS and adopt $A_V=3.1\,E(B-V)$ \citep{zucker_deep_2025}. For each cluster, we query the dust maps at the position and distance of every member star. If all member queries return valid estimates, we adopt a Gaussian $A_V$ prior centered on the median member-level estimate, with $\sigma_{A_V}=0.10$ mag and the physical constraint $A_V\geq0$. Valid dust-map estimates are available for 98{,}000 of the 98{,}014 member stars. The remaining 14 belong to one nearby cluster that falls partly outside the available map coverage; for this cluster, the resulting prior requires only $A_V\geq0$. We adopt the posterior modes of $A_V$, \texttt{skewness}, and \texttt{scale}, with the $A_V$ uncertainty computed from its 68\% highest-density interval.

For every cluster, we identify separated components in the posterior distribution of $\log_{10}(\mathrm{age/yr})$ and select the component with the largest integrated posterior probability. We adopt the median age and conditional 68\% highest-density interval of that component. Of the 2508 fits, 1168 (47\%) contain multiple retained components; within the 2.5 kpc cube sample, these clusters tend to have lower inferred initial masses and fewer members than clusters with unimodal age posteriors.

We estimate cluster masses from the same fitted PARSEC models. For each posterior draw, we assign a mass to every member star by finding the nearest point on the fitted isochrone in color--magnitude space and taking the stellar mass associated with that isochrone point. We convert the apparent-magnitude interval $12 \leq G \leq 17$ to a cluster-specific stellar-mass interval using the cluster distance and the age and extinction of each posterior draw. We then compare the observed member star mass function within this interval with synthetic Kroupa IMF clusters spanning total masses from 10 to 5000 solar masses. Thus, the stellar-mass range used in the fit varies among clusters. The adopted present-day mass is the posterior median of the best-fitting total masses, and the uncertainty is the 16th--84th percentile range.

Our present-day cluster masses omit stars dispersed into the field through internal relaxation and external perturbations from the Galaxy and molecular clouds \citep{lamers_analytical_2005,lamers_clusters_2006,gieles_star_2006}. The depletion of older and lower-mass clusters is apparent in Gaia cluster censuses \citep{anders_star_2021,hunt_improving_2024}. We approximate this lost mass by inferring the initial mass, $M_i$, through numerical inversion of Equation~(2) from \citet{almeida_open_2024},
\begin{equation}
M(t)=M_i-\frac{t}{t_4\gamma}M_i^{1-\gamma}(10^4\,M_\odot)^\gamma,
\end{equation}
with $t_4=2.9$ Gyr and $\gamma=0.7$. We propagate the present-day mass uncertainty through the same inversion and use the inferred initial mass and uncertainty for the star formation history and IMF sampling. Combining the bulk velocity sample with the successful \texttt{Chronos} age/mass fits and requiring an adopted \texttt{Chronos} age $<100$ Myr gives the 2.5 kpc cube sample of 568 clusters used in the reconstruction (Table~\ref{tab:sample}). Applying the 500 pc cut defined in \S\ref{sec:data_catalog} gives the 1 kpc cube sample of 312 clusters. Figure~\ref{fig:age_mass_comparison} compares the resulting ages and masses to the original HR23 catalog values.

\subsection{Cluster orbits in a Galactic potential}\label{sec:orbits}

We trace each cluster backward in time with the \texttt{galpy} package \citep{bovy_galpy_2015}, following the strategy of \citet{swiggum_most_2024}. The initial coordinates are the present-day centroid position and velocity $(X,Y,Z,U,V,W)$ for each cluster. We integrate each orbit 100 Myr into the past and sample it on a 1 Myr grid. We adopt the \texttt{MWPotential2014} axisymmetric Milky Way potential, a solar Galactocentric radius of $R_0 = 8.122$ kpc \citep{gravity_collaboration_detection_2018}, a local circular speed of $V_{\rm circ} = 236~{\rm km~s^{-1}}$ \citep{nitschai_first_2020}, and a solar height above the Galactic midplane of $z_\odot = 20.8$ pc \citep{bennett_vertical_2019}. We correct velocities to the local standard of rest using the solar peculiar motion $(U_\odot, V_\odot, W_\odot) = (11.1, 12.24, 7.25)~{\rm km~s^{-1}}$ from \citet{schonrich_local_2010}.

Uncertainties are propagated with Monte Carlo orbit sampling. For each cluster, we generate 100 phase-space realizations by drawing the present-day centroid position and velocity from Gaussian distributions centered on the measured $(X,Y,Z,U,V,W)$ values, with widths set by their reported uncertainties. We hold the adopted Galactic parameters and Solar peculiar motion fixed since our analysis depends primarily on the relative orbits between clusters, which change little when these parameters are varied \citep{swiggum_most_2024}. This time window balances the goal of following recent cluster motions against the increasing uncertainty at earlier times: a velocity uncertainty of $1~{\rm km~s^{-1}}$ accumulates to roughly 100 pc over 100 Myr.

Additional uncertainty comes from the use of an axisymmetric potential, since time-dependent non-axisymmetric structure can perturb stellar orbits. Simulations by \citet{arunima_their_2025} show that spiral arms and giant molecular clouds can deflect stellar orbits over $\lesssim100$ Myr. However, \citet{arunima_evolution_2026} find that stars born close together can retain correlated orbital changes for up to 0.5 Gyr. Although our reconstruction uses the absolute cluster orbits, the main results depend most strongly on the relative orbits of recently co-born clusters, even if the absolute positions of the clusters over the traceback are more uncertain.

We express the integrated cluster tracebacks in a local frame that co-moves and co-rotates with the LSR. The frame origin follows a circular orbit at $R_0$, and the $X,Y$ basis rotates with angular speed $\Omega_0 = V_{\rm circ}/R_0$, so $X$ remains directed toward the Galactic center and $Y$ remains tangential to Galactic rotation at each lookback time. We use this frame when overlaying the recent ccSN reconstruction on the present-day dust map in Figure~\ref{fig:recent_map}. For the broader traceback in Figure~\ref{fig:timeline}, we show the reconstruction in heliocentric Galactic Cartesian $X,Y$ coordinates.

\subsection{IMF sampling and supernovae locations}\label{sec:sn_generation}

For each cluster, we use the inferred initial cluster mass and its uncertainty together with a Kroupa IMF \citep{kroupa_variation_2001} to generate an ensemble of possible stellar populations. We sample the IMF over $0.03$--$120\,M_\odot$. We draw 100 IMF realizations per cluster and retain sampled stars with initial masses $M_{\rm init}\geq8\,M_\odot$ as candidate ccSN progenitors. For each realization, we follow these massive stars from cluster formation to their adopted death times, allowing us to reconstruct their ionizing and wind output before any resulting ccSN. Repeating this procedure captures the stochastic uncertainty in the number and masses of massive stars expected for a cluster of a given mass, which is especially important for low- and intermediate-mass systems where the upper IMF is only sparsely populated.

\begin{figure}[!t]
    \centering
    \includegraphics[width=0.90\columnwidth]{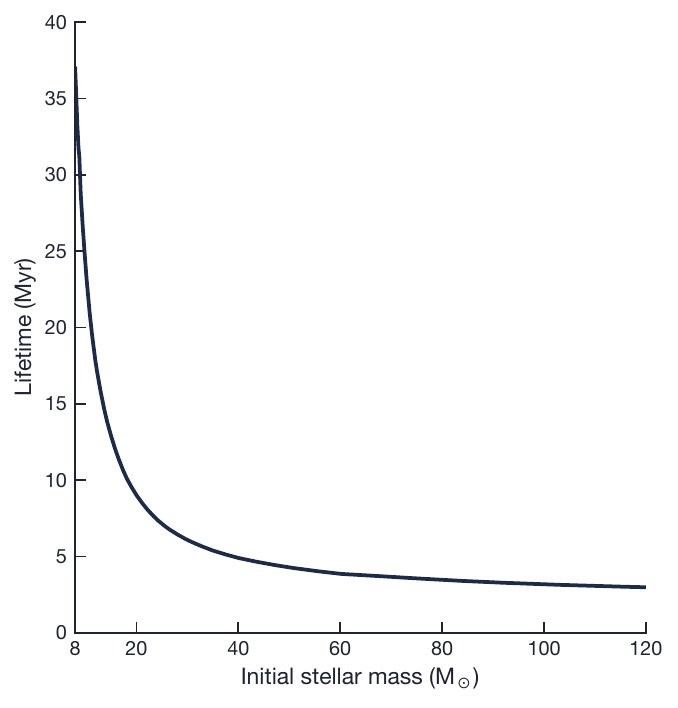}
    \caption{Stellar lifetime as a function of initial mass for the non-rotating, solar-metallicity PARSEC v2.0 tracks adopted in the ccSN reconstruction, with $Z=0.014$ and $Y=0.273$ \citep{bressan_span_2012,costa_evolutionary_2025}. The curve shows the single-star relation over the $8$--$120\,M_\odot$ progenitor range; the interacting-binary lifetime adjustment described in the text is applied during population sampling.}
    \label{fig:stellar_lifetime}
\end{figure}

We assign each sampled progenitor a stellar lifetime by interpolating the non-rotating, solar-metallicity PARSEC v2.0 tracks, with a metal mass fraction $Z=0.014$ and helium mass fraction $Y=0.273$ \citep{bressan_span_2012,costa_evolutionary_2025}. Figure~\ref{fig:stellar_lifetime} shows the adopted single-star lifetime as a function of initial mass over the $8$--$120\,M_\odot$ progenitor range. We assume that each cluster is coeval, so the present-day cluster age sets the time elapsed since the birth of its stellar population. Before selecting past explosions, we include simplified prescriptions for interacting binaries, direct collapse, and ejected massive stars. Motivated by the high incidence of binary interaction among massive stars and its effect on the ccSN delay-time distribution \citep{sana_binary_2012,zapartas_delay-time_2017}, we shorten the lifetimes of a random 50\% of stars with initial masses $\geq15\,M_\odot$ by a half-normal fractional offset with $\sigma=0.15$. To represent the possibility of failed explosions at high initial masses, we remove 20\% of progenitors above $25\,M_\odot$ \citep{smartt_progenitors_2009,sukhbold_core-collapse_2016}. We also remove 10\% of already-exploded progenitors to account approximately for runaway or walkaway stars whose explosions would be displaced from the parent-cluster traceback \citep{carretero-castrillo_galactic_2023,renzo_massive_2019}. These choices are simplified approximations to effects that we do not model in full. A sampled progenitor contributes a past core-collapse supernova if its lifetime is shorter than the present-day cluster age of its host cluster; otherwise it remains alive at the present epoch. For progenitors that do explode in the past, we define the explosion lookback time as the difference between the cluster age and the stellar lifetime.

We assign each supernova a 3D location by evaluating the traced cluster orbit at that explosion lookback time and then sampling a intra-cluster offset. Specifically, each event is first placed at the position of its host cluster along the backward-integrated orbit at the epoch when the progenitor reaches the end of its life. We then draw an isotropic random offset uniformly in volume within the HR23 total cluster radius, \texttt{radius\_total\_pc}, and add that offset to the orbit position. This preserves the traced bulk motion of each parent cluster while preventing multiple supernovae from the same cluster from being artificially concentrated at a single point. Repeating this procedure over all IMF realizations and over the Monte Carlo orbit realizations described in \S\ref{sec:orbits} yields an ensemble of explosion times and 3D positions for each cluster. These events are then combined across the 2.5 kpc cube sample to construct the supernova maps, time-dependent rates, regional counts, and clustering statistics presented in \S\ref{sec:results}.

Our implementation adopts several simplifying assumptions. We treat each cluster as a single coeval population, use the present-day cataloged total radius as a time-independent spatial scale, and do not model full binary population synthesis, explicit runaway trajectories, or cluster expansion nor rotation. The Almeida et al. correction accounts for secular cluster mass loss but not early gas expulsion or stars born outside the surviving cluster population. The resulting supernova map should therefore be interpreted as the clustered component of the recent core-collapse supernova history that is recoverable from the surviving young cluster population.

\subsection{Clustering tests and randomized comparisons}\label{sec:clustering_methods}

To test whether the reconstructed ccSNe are more spatially clustered than expected from randomized cluster locations, we construct two null comparisons from an ensemble of 100 randomized-cluster realizations. Both the reconstruction and the randomized-cluster comparison use the same 312 clusters in the 1 kpc cube sample. For each randomized counterpart, we retain the observed cluster age, inferred initial mass, and radius while drawing $X$ and $Y$ uniformly within $\pm500$ pc. We jointly permute the observed $Z$ coordinate and residual velocity vector after subtracting the local circular velocity, then add that residual to the circular velocity predicted by \texttt{MWPotential2014} at the randomized position. This preserves the cluster population used to generate the ccSNe and changes only its present-day phase-space arrangement.

The randomized-cluster comparison generates ccSNe from the retained ages and masses, so multiple explosions remain associated with the same randomized host. The randomized-location comparison instead keeps the reconstructed ccSN times fixed and assigns each event to an orbit from the same 312-cluster randomized ensemble at the corresponding lookback time. Within each 10 Myr time bin, we sample the randomized carrier clusters without replacement. This disperses individual ccSNe without introducing repeated random locations within the bin and tests the total spatial clustering after controlling for the time dependence of the ccSN rate.

We quantify clustering with the median nearest-neighbor distance in the Galactic $X$--$Y$ plane, calculated in independent, non-overlapping 10 Myr bins. We downsample all three catalogs without replacement to the common available event count in each paired realization and time bin, preventing residual differences in event density from setting the nearest-neighbor ratio. For the randomized-cluster comparison, we calculate the statistic using all ccSN pairs and separately using only pairs from different hosts. The latter isolates clustering among separate cluster hosts. For the randomized-location comparison, we use the nearest ccSN regardless of host and therefore measure the total clustering, including repeated explosions from the same cluster. We report $d_{\rm rand}/d_{\rm data}$, where values above unity mean that the reconstructed ccSNe are more clustered than the randomized comparison.

\subsection{Calculation of star formation, supernova rates, and feedback power}\label{sec:rate_methods}

We compute the cluster-based star formation history by treating each cluster as a single burst with total stellar mass given by the inferred initial mass in \S\ref{sec:ages}. Each cluster contributes its mass at its birth time given by its age and smoothed by a Gaussian kernel whose width is set by the cluster age uncertainty, with a minimum width of 0.5 Myr. Summing these kernels gives the star formation rate (SFR) as a function of lookback time, i.e. the star formation history.

For the ccSN history, we construct a separate rate curve for each of the 100 IMF realizations. Each realization includes draws from the inferred cluster age and initial-mass distributions, propagating these uncertainties together with stochastic IMF sampling and the population prescriptions described in \S\ref{sec:sn_generation}. Each reconstructed ccSN contributes a unit-normalized Gaussian kernel with $\sigma=3$ Myr, and summing these kernels gives the ccSN rate in Myr$^{-1}$. This fixed kernel sets the temporal smoothing scale; the cluster age uncertainties are propagated through the age draws. We report the mean and 16th--84th percentile range across the 100 realizations.

Using the same IMF realizations, we calculate the time-dependent power produced by the retained $M_{\rm init}\geq8\,M_\odot$ candidate ccSN progenitors from cluster formation to their adopted death times. We estimate the ionizing luminosity by integrating a blackbody with the PARSEC luminosity and effective temperature above 13.6 eV. We calculate the stellar-wind mechanical power as
\begin{equation}
L_{\rm wind}=\frac{1}{2}\dot{M}v_{\infty}^{2},
\end{equation}
using the PARSEC mass-loss rate and $v_{\infty}=C(T_{\rm eff})v_{\rm esc}$, where $C=v_{\infty}/v_{\rm esc}$ is the ratio of the terminal wind speed to the stellar escape speed. We adopt $C=1.0$, 1.4, and 2.65 for $T_{\rm eff}\leq10{,}000$ K, $10{,}000<T_{\rm eff}<21{,}000$ K, and $T_{\rm eff}\geq21{,}000$ K, respectively, following \citet{kudritzki_winds_2000}. Although selected by initial mass, these progenitors can enter cooler evolutionary phases, so we evaluate $C$ from their effective temperature at each age. We smooth the ionizing and wind histories in each realization with a Gaussian kernel with $\sigma=0.5$ Myr. We convert the smoothed ccSN rate to mechanical power by assigning $10^{51}$ erg to each explosion. We define the source-level massive-star feedback power as the sum of the ionizing luminosity, stellar-wind mechanical power, and ccSN mechanical power. It does not account for photon escape, radiative losses, or coupling to the surrounding gas.

Figure~\ref{fig:local_history} shows the star formation, ccSN, and feedback power histories for the 1 kpc cube sample. The left column includes all activity associated with clusters selected in the present-day cube. The right column retains only activity located inside the traced cube at each lookback time and reports the corresponding surface densities.

We also convert each local ccSN surface density in Table~\ref{tab:rates_clusters_ob} to a Milky Way-equivalent rate. We follow the local-to-Galactic extrapolation of \citet{quintana_census_2025}, which adopts the axisymmetric exponential-disk and central-hole model of \citet{reed_new_2005}. To reproduce this calculation, we retain the values adopted by \citet{quintana_census_2025}: $R_0=8.5$ kpc, a disk scale length $H=0.28R_0$, and a hole radius $R_{\rm hole}=0.5R_0$. We normalize this radial profile to the local surface density. Integrating the profile gives an effective Galactic area of $A_{\rm eff}=720~\mathrm{kpc}^{2}$, so $R_{\rm MW}=\Sigma_{\rm ccSNe}A_{\rm eff}$ after converting from Myr$^{-1}$ to century$^{-1}$. The cluster-based rate is $R_{\rm MW}=0.55\pm0.03$ century$^{-1}$; a uniform disk with a 15 kpc radius gives 0.54 century$^{-1}$. The quoted uncertainty propagates only the formal uncertainty in the local cluster-based rate. This extrapolation assumes that the Solar Neighborhood rate is representative at the Solar radius and does not correct for ccSN progenitors missing from the input sample.

\begin{figure*}[!t]
    \centering
    \includegraphics[width=\textwidth]{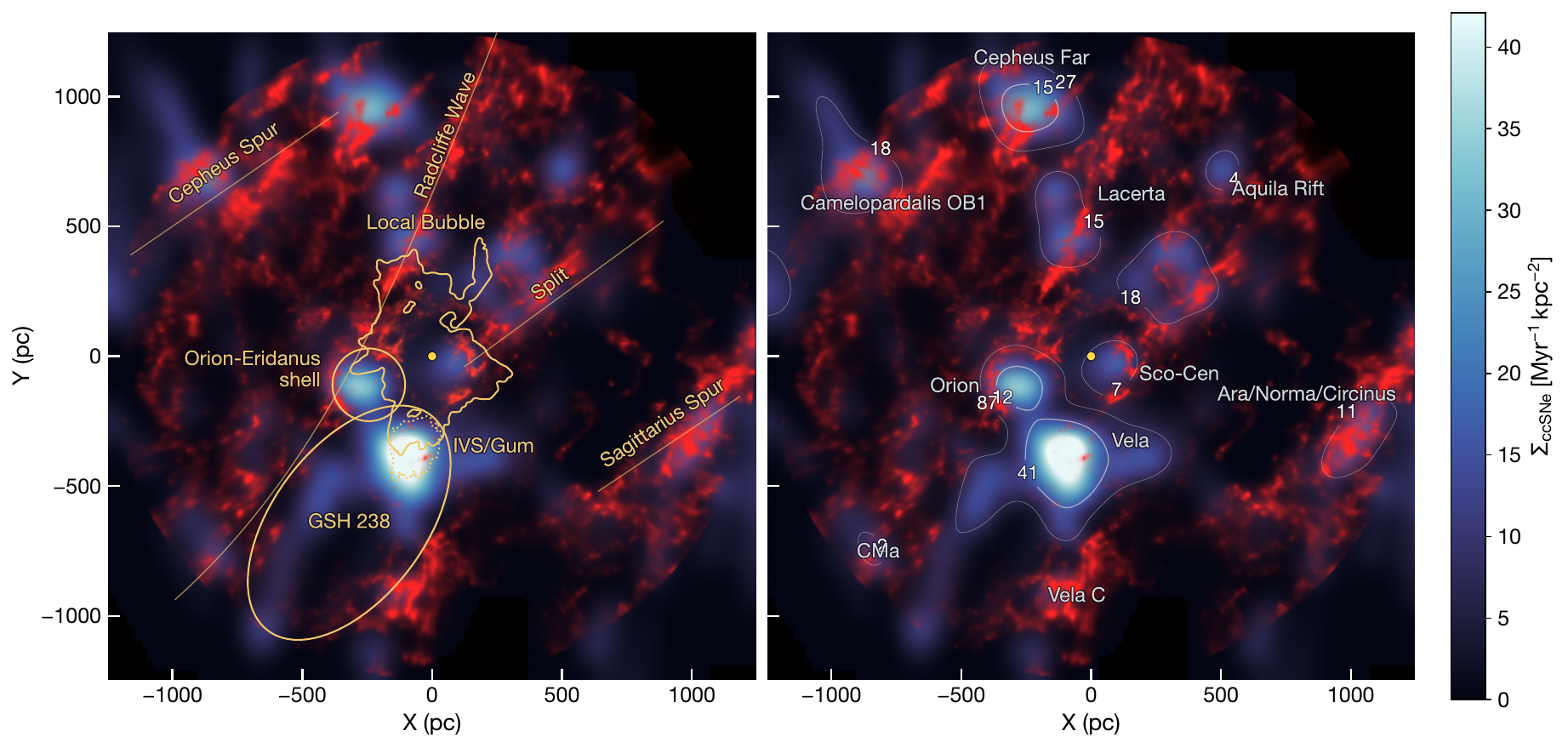}
    \caption{Recent nearby core-collapse supernova rate surface density, \(\Sigma_{\mathrm{ccSNe}}\), in the local Galactic $X$--$Y$ plane, integrated over the last 15 Myr and over $|Z| < 500$ pc. Both panels show the present-day dust distribution from the parsec-scale 3D map of \citet{edenhofer_parsec-scale_2024} in red and the reconstructed \(\Sigma_{\mathrm{ccSNe}}\) map in blue. In the left panel, gold lines and labels mark large-scale dust spines or cloud complexes (the Radcliffe Wave, Cepheus Spur, Split, and Sagittarius Spur) and shell or bubble models (the Local Bubble, Orion--Eridanus shell, IVS/Gum, and GSH 238). The dotted IVS/Gum outline is the top-down projection of its 3D outer shell boundary. Grey contours in the right panel are drawn automatically from the smoothed ccSN map at $\Sigma_{\mathrm{ccSNe}}=7$ and $17~{\rm Myr}^{-1}~{\rm kpc}^{-2}$; the white numbers give the rounded mean number of reconstructed ccSNe enclosed by each closed contour loop across IMF realizations. Grey text labels identify nearby regional enhancements and are placed by hand only for orientation. The yellow point marks the present-day Sun. The dust map is shown only for present-day context and should not be interpreted as a coeval gas distribution at the epoch of the explosions. An interactive version is available at \url{https://cswigg.github.io/cam_website/oviz_figures/oviz_supernovae.html}.}
    \label{fig:recent_map}
\end{figure*}

For map visualizations, we convert the discrete reconstructed explosions into \(\Sigma_{\mathrm{ccSNe}}\) maps by smoothing each event in the relevant $X$--$Y$ plane with a Gaussian spatial kernel of standard deviation $\sigma=50$ pc and dividing by the adopted time window. Figure~\ref{fig:recent_map} uses the co-moving, co-rotating LSR frame described in \S\ref{sec:orbits}, allowing comparison with the present-day dust map. We integrate over the last 15 Myr and include explosions with $|Z|<500$ pc. Figure~\ref{fig:timeline} shows the broader traceback in heliocentric Galactic Cartesian $X,Y$ coordinates using 10 Myr windows and the same 50 pc kernel.

We also use the two observed OB star catalogs described in \S\ref{sec:data_ob_catalogs} as a separate normalization check. We compare Q25 and ALS III with the synthetic massive star population predicted by the 1 kpc cube sample, $|X|, |Y|, |Z| \leq 500$ pc. The projected area is therefore $A=1~{\rm kpc}^{2}$. For the cluster reconstruction, $N_{>8\,M_{\odot}}$ is the median number of synthetic stars with initial mass $m \geq 8\,M_{\odot}$ and future death times, $t_{\rm death}>0$, across IMF realizations.

The OB star rate conversion uses the same Kroupa IMF and PARSEC stellar lifetimes adopted for the cluster sampling. For a constant recent star formation rate, the expected number of living stars above a mass $m_0$ is
\begin{equation}
N_{\rm alive}(>m_0) =
\Sigma_{\mathrm{SFR}}\,A\,C_{\rm alive}(>m_0),
\end{equation}
where
\begin{equation}
C_{\rm alive}(>m_0) =
\int_{m_0}^{m_{\rm max}} \tau(m)\,\phi(m)\,dm .
\end{equation}
Here $\tau(m)$ is the stellar lifetime and $\phi(m)$ is the Kroupa IMF normalized per unit stellar mass formed. For the adopted tracks, an $8\,M_{\odot}$ star has a lifetime of 36.96 Myr and $C_{\rm alive}(>8\,M_{\odot}) = 0.17713~{\rm Myr}\,M_{\odot}^{-1}$.

We therefore calculate
\begin{equation}
\Sigma_{\mathrm{SFR}} =
\frac{N_{\rm alive}(>m_0)}
{A\,C_{\rm alive}(>m_0)}
\end{equation}
and
\begin{equation}
\Sigma_{\mathrm{ccSNe}} =
\eta_{\rm ccSN}(>8\,M_{\odot})\,\Sigma_{\mathrm{SFR}},
\end{equation}
where $\eta_{\rm ccSN}(>8\,M_{\odot}) = 0.01054~{\rm SN}\,M_{\odot}^{-1}$. We apply this conversion to the Q25 $M \geq 8\,M_{\odot}$ sample and to the ALS III $\texttt{Cat}=\texttt{M}$ sample. We then use the resulting OB-star $\Sigma_{\rm ccSNe}$ values to calculate Milky Way-equivalent ccSN rates with the same Galactic extrapolation described above.

The OB-star rate estimates assume that the Q25 $M\geq8\,M_{\odot}$ and ALS III $\texttt{Cat}=\texttt{M}$ selections trace the same population of massive stars. ALS III uses the 20 kK extinction track as a photometric proxy for this mass threshold. Both selections rely primarily on photometry and can be affected by degeneracies among temperature, extinction, and luminosity, variations in the extinction law, stellar evolution, and measurement errors \citep{pantaleoni_gonzalez_alma_2025}. Because we assign individual masses to cluster members from their best-fitting PARSEC isochrones in \S\ref{sec:ages}, we can use these masses to test whether the two catalogs include lower-mass stars or miss stars above $8\,M_{\odot}$. We crossmatch the cluster members to Q25 and ALS III using their \textit{Gaia} DR3 source IDs. For Q25, we consider stars in the 1 kpc cube both above and below the $8\,M_{\odot}$ catalog threshold. For ALS III, we consider stars in the same volume both inside and outside the $\texttt{Cat}=\texttt{M}$ selection. We define the contamination fraction as the fraction of stars inside the primary catalog selection that the reference diagnostic classifies below the massive-star threshold. We define the selection incompleteness fraction as the fraction of reference massive stars excluded by the primary catalog selection. We use these fractions to calculate the alternative rate normalization implied by the cluster-mass comparison. These fractions are estimated only from crossmatched cluster members and may not represent the full catalog, particularly when few matched stars fall outside the primary selection. Individual stellar masses of the clusters remain uncertain because they depend on the adopted isochrone models. Therefore, this analysis is meant only as a comparison. 

Spectral classifications from high-resolution spectra provide a much closer empirical reference as they constrain stellar temperature directly from spectroscopy rather than broadband photometry. We consider spectral classifications from the working GOSC/ALS III database (M. Pantaleoni Gonz\'alez 2026, private communication). Where available, we also crossmatch these classifications to Q25 within the same 1 kpc cube. For both catalogs, we compare spectroscopically identified massive OB and low-mass ($<8\,M_{\odot}$) B stars both inside and outside the primary massive-star selection. This allows us to estimate the contamination and selection incompleteness fractions and calculate a second rate normalization implied by the spectroscopic comparison. The complete source-level classifications and crossmatches are not yet public, and the spectroscopic targets were not selected randomly. For example, preferentially targeting stars in known OB associations would overrepresent regions already expected to contain massive stars, so the measured contamination and selection incompleteness fractions may not apply to the full field population.

The resulting OB star numbers and inferred star formation and supernovae rates are reported in Table~\ref{tab:rates_clusters_ob}, underneath the corresponding cluster-based values.

\begin{deluxetable*}{lccccc}[htb!]
\tabletypesize{\footnotesize}
\tablewidth{\textwidth}
\tablecaption{Local SFR and ccSN surface densities, ccSN energy released over 40 Myr, and Milky Way-equivalent ccSN rates inferred from clusters and OB stars in the 1 kpc cube sample\label{tab:rates_clusters_ob}}
\tablecolumns{6}
\tablehead{
\colhead{Sample} &
\colhead{$\Sigma_{\mathrm{SFR}}$} &
\colhead{$\Sigma_{\mathrm{ccSNe}}$} &
\colhead{$E_{\mathrm{ccSNe}}$} &
\colhead{$N_{>8\,\mathrm{M}_{\odot}}$} &
\colhead{$R_{\rm MW}$} \\
\colhead{} &
\colhead{\footnotesize{$\mathrm{M}_{\odot}\,\mathrm{Myr}^{-1}\,\mathrm{kpc}^{-2}$}} &
\colhead{\footnotesize{$\mathrm{Myr}^{-1}\,\mathrm{kpc}^{-2}$}} &
\colhead{\footnotesize{$10^{53}\,\mathrm{erg}$}} &
\colhead{} &
\colhead{\footnotesize{century$^{-1}$}}
}
\startdata
\cutinhead{\textbf{Clusters}}
\hspace{0.6em}1 kpc cube sample & $823^{+19}_{-19}$ & $7.7^{+0.4}_{-0.4}$ & $3.1^{+0.2}_{-0.2}$ & $111^{+8}_{-8}$ & $0.55^{+0.03}_{-0.03}$ \\
\cutinhead{\textbf{OB stars}}
\hspace{0.6em}Q25 & 900 & 10 & 4.0 & 164 & 0.71 \\
\hspace{0.6em}ALS III & 3700 & 39 & 15.6 & 648 & 2.78 \\
\cutinhead{\textbf{OB stars (Isochrone mass comparison)}}
\hspace{0.6em}Q25 & 1200 & 13 & 5.2 & 219 & 0.94 \\
\hspace{0.6em}ALS III & 800 & 8 & 3.2 & 133 & 0.57 \\
\cutinhead{\textbf{OB stars (Spectral class comparison)}}
\hspace{0.6em}Q25 & 1700 & 18 & 7.2 & 293 & 1.26 \\
\hspace{0.6em}ALS III & 4400 & 47 & 18.8 & 781 & 3.35 \\
\enddata
\tablecomments{\scriptsize The cluster $\Sigma_{\rm SFR}$ and $\Sigma_{\rm ccSNe}$ values are averages over the past 40 Myr. Cluster uncertainties give formal 16th--84th percentile intervals from age and mass draws for the SFR and 100 IMF realizations for the ccSN rate and count. The OB-star rates are steady-state estimates from present-day massive-star counts. OB-star $\Sigma_{\rm SFR}$ and $\Sigma_{\rm ccSNe}$ values are rounded to the nearest $100$ and $1$, respectively. The $E_{\mathrm{ccSNe}}$ column gives the nominal ccSN energy released over 40 Myr within the $1~\mathrm{kpc}^{2}$ comparison area, assuming $10^{51}$ erg per ccSN and no coupling correction. For the OB-star rows, these values give the energy implied over 40 Myr if the steady-state rate remains constant. Ratios quoted in the text and $R_{\rm MW}$ values are calculated from the unrounded rates. We calculate $R_{\rm MW}$ by applying the same $A_{\rm eff}=720~\mathrm{kpc}^{2}$ disk-plus-hole extrapolation to every row \citep{reed_new_2005,quintana_census_2025}. The $R_{\rm MW}$ column does not include uncertainty in the Galactic profile. Comparison rows report the alternative normalizations implied by the two overlap samples; they do not define revised catalogs.}
\end{deluxetable*}

\section{Results}\label{sec:results}

\subsection{Recent nearby ccSNe and present-day cavities}\label{sec:results_recent_map_draft}

Figure~\ref{fig:recent_map} shows the recent reconstructed \(\Sigma_{\mathrm{ccSNe}}\) in the Solar Neighborhood over the last 15 Myr. The reconstruction is shown in the local Galactic $X$--$Y$ plane, integrated over $|Z| < 500$ pc, and overlaid on the present-day dust distribution from \citet{edenhofer_parsec-scale_2024}. The dust is included as present-day ISM context for the reconstructed ccSNe. Several recent ccSN enhancements can be seen to occupy cavities in the present-day dust map.

Several of the strongest recent enhancements lie within known present-day shells and cavities. We quantify this correspondence by counting reconstructed ccSNe inside the shell volumes or projected footprints shown in the left panel of Figure~\ref{fig:recent_map}, using time windows chosen to approximate the relevant feedback history of each structure. For the Local Bubble, we use the 3D shell model of \citet{oneill_local_2024} and integrate over 15 Myr, close to the $\sim14$ Myr expansion timescale found by \citet{zucker_star_2022}. For IVS/Gum, we count ccSNe inside the 3D outer shell boundary from \citet{gao_origin_2025} and integrate over the last 3 Myr, matching the recent-SN window used in their feedback budget. Figure~\ref{fig:recent_map} shows the top-down projection of this boundary. For Orion--Eridanus and GSH 238+00+09, which do not yet have comparable 3D shell models, we use projected top-down rings drawn by eye from the present-day dust morphology. We integrate Orion--Eridanus over 15 Myr and GSH 238+00+09 over 30 Myr. The resulting counts are reported in Table~\ref{tab:shell_counts}.

The right panel of Figure~\ref{fig:recent_map} shows the same recent ccSN reconstruction with contours drawn at \(\Sigma_{\mathrm{ccSNe}}=7\) and \(17~{\rm Myr}^{-1}~{\rm kpc}^{-2}\). These two levels were chosen such that the resulting contours enclose the main visually prominent enhancements in the recent map. The enclosed regions include Vela, Cepheus Far, Orion, Sco--Cen, Camelopardalis OB1, Lacerta, Aquila Rift, Ara/Norma/Circinus, and CMa. We also report the reconstructed ccSN counts inside these contours in Table~\ref{tab:contour_counts}. These counts depend on the adopted contour levels and should be interpreted as approximate summaries of the main recent enhancements.

\begin{table*}[!tp]
\centering
\footnotesize
\setlength{\tabcolsep}{3.5pt}
\renewcommand{\arraystretch}{1.10}
\caption{Reconstructed core-collapse supernova counts in the physical bubble and shell regions shown in the context panel of Figure~\ref{fig:recent_map}. The Local Bubble and IVS/Gum rows use their 3D model volumes. Orion--Eridanus and GSH 238 use the displayed projected aperture and ellipse. Counts are means across IMF realizations; uncertainties give the 16th--84th percentile interval, rounded to the nearest event. The listed clusters are ordered by mean reconstructed contribution in each region.\label{tab:shell_counts}}
\begin{tabular}{lccl}
\hline
\hline
Feature & Time window & $N_{\rm ccSNe}$ & Largest contributing clusters \\
& (Myr ago) & & \\
\hline
Local Bubble & 0--15 & $9^{+3}_{-3}$ &
\begin{minipage}[t]{0.48\textwidth}\raggedright
V1062 Sco (UPK 640); IC 2602; $\nu$ Cen (HSC 2636); HSC 2468; HSC 976
\end{minipage} \\
Orion--Eridanus shell & 0--15 & $17^{+5}_{-4}$ &
\begin{minipage}[t]{0.48\textwidth}\raggedright
Collinder 69 / $\lambda$ Orionis; ZHBJZ 1; NGC 2232; UPK 422; NGC 1980
\end{minipage} \\
IVS/Gum & 0--3 & $2^{+2}_{-1}$ &
\begin{minipage}[t]{0.48\textwidth}\raggedright
OC 0470; Pozzo 1; NGC 2451B; NGC 2547; OC 0450
\end{minipage} \\
GSH 238 & 0--30 & $132^{+11}_{-11}$ &
\begin{minipage}[t]{0.48\textwidth}\raggedright
Haffner 13; Alessi 34; Theia 2267; NGC 2547; NGC 2451B
\end{minipage} \\
\hline
\end{tabular}

\vspace{1.0em}

\caption{Reconstructed core-collapse supernova counts enclosed by the Figure~\ref{fig:recent_map} \(\Sigma_{\mathrm{ccSNe}}\) contours. All rows use the 0--15 Myr window and $|Z|<500$ pc. Counts are means across IMF realizations; uncertainties give the 16th--84th percentile interval, rounded to the nearest event. The listed clusters are ordered by mean reconstructed contribution inside each contour.\label{tab:contour_counts}}
\begin{tabular}{lcl}
\hline
\hline
Feature & $N_{\rm ccSNe}$ & Largest contributing clusters \\
\hline
\multicolumn{3}{l}{$17~{\rm Myr}^{-1}\,{\rm kpc}^{-2}$ contour} \\
\hline
Vela & $41^{+7}_{-7}$ &
\begin{minipage}[t]{0.55\textwidth}\raggedright
Alessi 34; Haffner 13; NGC 2547; NGC 2451B; LISC-III 3668
\end{minipage} \\
Orion & $12^{+4}_{-3}$ &
\begin{minipage}[t]{0.55\textwidth}\raggedright
ZHBJZ 1; NGC 2232; UPK 422; Collinder 69; NGC 1980
\end{minipage} \\
Cepheus Far & $15^{+4}_{-4}$ &
\begin{minipage}[t]{0.55\textwidth}\raggedright
OC 0185; ASCC 114; UPK 172; NGC 7160; Pismis-Moreno 1
\end{minipage} \\
\hline
\multicolumn{3}{l}{$7~{\rm Myr}^{-1}\,{\rm kpc}^{-2}$ contour} \\
\hline
Vela--Orion--Sco--Cen complex & $87^{+9}_{-9}$ &
\begin{minipage}[t]{0.55\textwidth}\raggedright
Alessi 34; Haffner 13; IC 2395; Trumpler 10; NGC 2547
\end{minipage} \\
Cepheus Far & $27^{+6}_{-6}$ &
\begin{minipage}[t]{0.55\textwidth}\raggedright
FSR 0398; ASCC 114; OC 0185; HSC 764; CWNU 136
\end{minipage} \\
Camelopardalis OB1 & $18^{+3}_{-3}$ &
\begin{minipage}[t]{0.55\textwidth}\raggedright
Theia 1722; NGC 1502; UBC 51; NGC 1444; Trumpler 3
\end{minipage} \\
Aquila Rift & $18^{+4}_{-5}$ &
\begin{minipage}[t]{0.55\textwidth}\raggedright
Theia 38; UBC 26; Melotte 186; Alessi 19; Stephenson 1
\end{minipage} \\
Lacerta & $15^{+4}_{-4}$ &
\begin{minipage}[t]{0.55\textwidth}\raggedright
UPK 166; UPK 168; RSG 8; ASCC 127; Theia 100
\end{minipage} \\
Aquila Rift secondary contour & $4^{+2}_{-1}$ &
\begin{minipage}[t]{0.55\textwidth}\raggedright
ASCC 107; Theia 316; OC 0077; UPK 62; Roslund 1
\end{minipage} \\
Ara/Norma/Circinus & $11^{+3}_{-4}$ &
\begin{minipage}[t]{0.55\textwidth}\raggedright
OC 0684; HSC 2911; BH 221; Theia 1645; NGC 6250
\end{minipage} \\
Sco--Cen & $7^{+2}_{-3}$ &
\begin{minipage}[t]{0.55\textwidth}\raggedright
V1062 Sco (UPK 640); IC 2602; $\sigma$ Sco (HSC 2907); $\nu$ Cen (HSC 2636); HSC 2468
\end{minipage} \\
CMa & $2^{+2}_{-1}$ &
\begin{minipage}[t]{0.55\textwidth}\raggedright
BDSB 96; OC 0377; vdBergh 92; Ruprecht 26; Theia 3581
\end{minipage} \\
\hline
\end{tabular}
\end{table*}

\subsection{The 50 Myr traceback map}\label{sec:results_traceback_draft}

\begin{figure*}[t!]
    \centering
    \includegraphics[width=\textwidth]{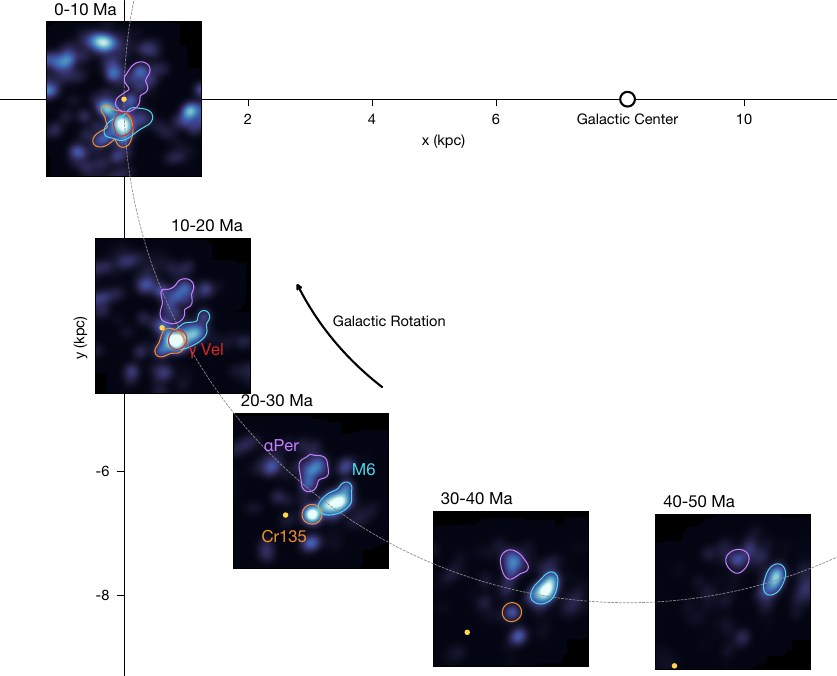}
    \caption{Time-resolved nearby core-collapse supernova rate surface density, \(\Sigma_{\mathrm{ccSNe}}\), in five 10 Myr lookback windows spanning 0--50 Myr ago, each integrated over $|Z| < 500$ pc. The inset panels are positioned along the backward-traced orbit of the Solar Neighborhood around the Galactic center. The traced position of the Sun at the center of each lookback window is marked in yellow in every panel. Solid colored contours mark the smoothed projected density of reconstructed ccSNe whose parent clusters belong to the Collinder 135 (Cr135), Messier 6 (M6), Alpha Persei ($\alpha$~Per), and Gamma Velorum ($\gamma$~Vel) families. An interactive version is available at \url{https://cswigg.github.io/cam_website/oviz_figures/oviz_supernovae.html}.}
    \label{fig:timeline}
\end{figure*}

Figure~\ref{fig:timeline} extends the recent view in Figure~\ref{fig:recent_map} across the past 50 Myr. The panels follow the backward-traced orbit of the local standard of rest (LSR) around the Galactic center. It shows the reconstructed \(\Sigma_{\mathrm{ccSNe}}\) in 10 Myr lookback windows over the same projected local Galactic volume. The youngest panel recovers the recent Orion, Sco--Cen, Vela, and Cepheus Far enhancements discussed above. Table~\ref{tab:traceback_history} summarizes the ccSN counts and main contributing clusters within the Figure~\ref{fig:timeline} traceback footprint in each 10 Myr interval.

At earlier times, the map is dominated by three major cluster family enhancements: Collinder 135 (Cr135), Messier 6 (M6), and Alpha Persei ($\alpha$~Per). These are the three cluster families identified by \citet{swiggum_most_2024}, and their dashed contours mark ccSNe whose parent clusters belong to each family.

The ccSN contribution from the families is clearest in the 20--50 Myr panels, where the Cr135, M6, and $\alpha$~Per enhancements are still largely separable. In particular, ccSNe associated with the Cr135 and M6 families can be seen merging from the 20--30 Myr panel to the 10--20 Myr panel. At recent times, the Sco--Cen enhancement is the youngest nearby component of the broader $\alpha$~Per family, while the present-day Vela region still contains ccSNe from Cr135 family clusters, as well as the smaller $\gamma$~Vel family uncovered by \citet{swiggum_most_2024}. Weaker substructures are also visible outside these families but fade at earlier times, possibly because older dissolved clusters are increasingly absent from the surviving-cluster reconstruction. Figure~\ref{fig:timeline} links the recent nearby ccSN map to older cluster family structure over the past 50 Myr.

\begin{table*}[!tp]
\centering
\small
\renewcommand{\arraystretch}{1.25}
\caption{Core-collapse supernova counts and representative contributors in the traceback panels of Figure~\ref{fig:timeline}. Counts are measured within the $x',y'=\pm1250$ pc traceback panel footprint and $|Z|<500$ pc. Uncertainties give the 16th--84th percentile interval, rounded to the nearest event.\label{tab:traceback_history}}
\begin{tabular}{lcll}
\hline
\hline
Age & $N_{\rm ccSNe}$ & Description & Representative contributors \\
(Myr ago) & & & \\
\hline
40--50 & $70^{+9}_{-9}$ &
\begin{minipage}[t]{0.24\textwidth}\raggedright
The M6 and $\alpha$~Per families begin producing ccSNe.
\end{minipage} &
\begin{minipage}[t]{0.43\textwidth}\raggedright
NGC 6405\tablenotemark{b}; Alessi 5\tablenotemark{b}; Theia 3397; BH 99\tablenotemark{b}; Trumpler 10\tablenotemark{b}; ASCC 58\tablenotemark{b}; Alessi 21; Melotte 20\tablenotemark{c}
\end{minipage} \\[6pt]
30--40 & $95^{+12}_{-10}$ &
\begin{minipage}[t]{0.24\textwidth}\raggedright
The M6 family remains the most prominent enhancement; Cr135 family ccSNe first appear.
\end{minipage} &
\begin{minipage}[t]{0.43\textwidth}\raggedright
Trumpler 10\tablenotemark{b}; NGC 6405\tablenotemark{b}; Theia 2267; BH 99\tablenotemark{b}; Alessi 5\tablenotemark{b}; Theia 3397; CWNU 1044\tablenotemark{b}; ASCC 58\tablenotemark{b}
\end{minipage} \\[6pt]
20--30 & $137^{+12}_{-13}$ &
\begin{minipage}[t]{0.24\textwidth}\raggedright
M6 and Cr135-family ccSNe dominate.
\end{minipage} &
\begin{minipage}[t]{0.43\textwidth}\raggedright
Trumpler 10\tablenotemark{b}; NGC 6405\tablenotemark{b}; Haffner 13\tablenotemark{b}; BH 164\tablenotemark{b}; Theia 2267; ASCC 114; NGC 2547\tablenotemark{a}; NGC 2451B\tablenotemark{a}
\end{minipage} \\[6pt]
10--20 & $154^{+12}_{-13}$ &
\begin{minipage}[t]{0.24\textwidth}\raggedright
The Cr135 and M6 enhancements merge, with the $\gamma$~Vel family beginning to contribute at their intersection.
\end{minipage} &
\begin{minipage}[t]{0.43\textwidth}\raggedright
Alessi 34\tablenotemark{a}; Trumpler 10\tablenotemark{b}; Trumpler 3; UBC 26\tablenotemark{c}; Haffner 13\tablenotemark{b}; BH 164\tablenotemark{b}; NGC 6405\tablenotemark{b}; Theia 2267
\end{minipage} \\[6pt]
0--10 & $228^{+13}_{-14}$ &
\begin{minipage}[t]{0.24\textwidth}\raggedright
The recent enhancements are distributed across several nearby regions.
\end{minipage} &
\begin{minipage}[t]{0.43\textwidth}\raggedright
Theia 1722; OC 0684; NGC 1502; IC 2395\tablenotemark{a}; Alessi 34\tablenotemark{a}; Theia 38\tablenotemark{c}; OC 0185; Haffner 13\tablenotemark{b}
\end{minipage} \\[2pt]
\hline
\end{tabular}
\tablecomments{Cluster Family assignments: \tablenotemark{a} Collinder 135 (Cr135); \tablenotemark{b} Messier 6 (M6); \tablenotemark{c} Alpha Persei ($\alpha$~Per); \tablenotemark{d} Gamma Velorum ($\gamma$~Vel).}
\end{table*}

\subsection{Star formation, ccSN rates, and massive-star feedback power}\label{sec:results_rates}

\begin{figure*}[!t]
    \centering
    \includegraphics[width=\textwidth]{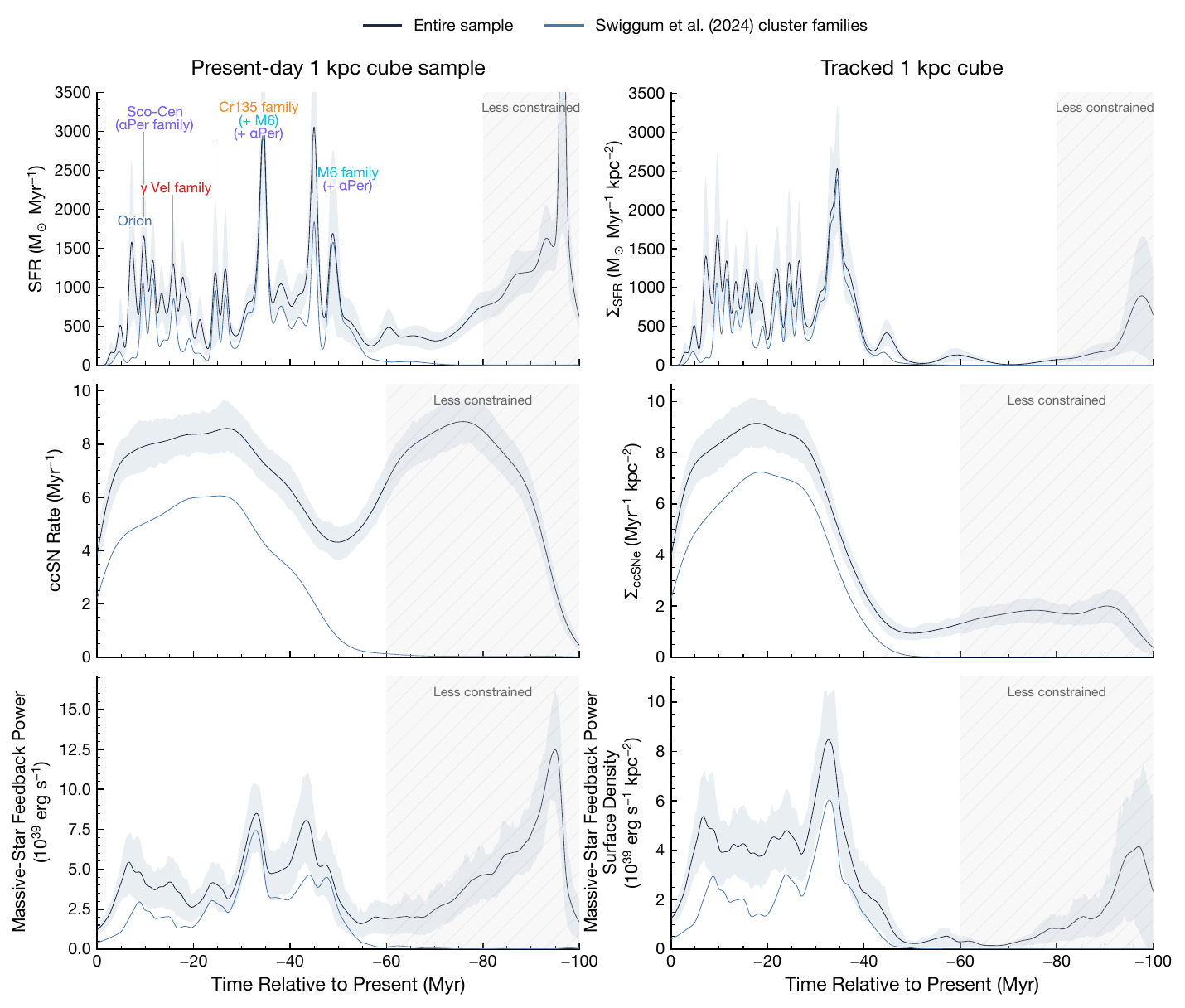}
    \caption{Star formation (top), ccSN (middle), and massive-star feedback power (bottom) histories for the 1 kpc cube sample. The left column includes all activity associated with clusters in the present-day cube; the right retains only activity inside the traced cube. Dark and blue curves show the full sample and cluster family contributions. Shading gives uncertainties, and hatching marks the less constrained intervals. Feedback power combines ionizing, wind, and ccSN contributions.}
    \label{fig:local_history}
\end{figure*}

Figure~\ref{fig:local_history} shows the SFR and ccSN rate histories, along with the massive-star feedback power history, for the 1 kpc cube sample. The left column follows clusters selected in the present-day cube and includes all activity associated with them. The right column shows the corresponding surface densities inside a co-moving, co-rotating 1 kpc cube, reported per $1~\mathrm{kpc}^{2}$ projected area. The bottom row combines ionizing luminosity, stellar-wind mechanical power, and ccSN mechanical power. Because clusters drift through the traced volume, the right column includes stellar radiation and winds only while the host lies inside the cube and includes ccSNe only when their reconstructed explosion positions lie inside it. At older lookback times, around 80--100 Myr, age and mass estimates become increasingly ambiguous near the ``isochrone blindspot'' identified by \citet{rottensteiner_empirical_2024}. We therefore mark 80--100 Myr in the star formation panels and 60--100 Myr in the ccSN and feedback power panels as less constrained. The earlier boundary for ccSNe and feedback power reflects contributions from clusters formed at larger lookback times.

In each panel of Figure~\ref{fig:local_history}, the blue curve shows the contribution from clusters associated with the Cr135, M6, $\gamma$~Vel, and $\alpha$~Per families identified by \citet{swiggum_most_2024}, while the dark curve shows the entire 1 kpc cube sample. The families contribute the majority of star formation and feedback, especially over the last $\sim40$ Myr.

The left column of Figure~\ref{fig:local_history} includes 312 present-day clusters within $|X|, |Y|, |Z|\leq500$ pc, of which 123 are assigned to the cluster families identified by \citet{swiggum_most_2024}. The right column uses the same 1 kpc cube sample but counts only the activity that falls inside the traced 1 kpc cube at each lookback time, reported per $1~\mathrm{kpc}^{2}$ projected area. Thus, the left column gives the history associated with the present-day 1 kpc cube sample, while the right column gives the part of that history that remains in the traced volume. The corresponding histories for the 2.5 kpc cube sample are shown in Figure~\ref{fig:xy1250_history}. Its SFR rises more sharply over the last 15 Myr because of additional activity beyond the 1 kpc cube, mainly toward Cepheus Far and Ara/Norma/Circinus.

The SFR rises after a dearth of activity before $\sim 60$ Myr ago, with several peaks corresponding to resolved bursts in the 1 kpc cube sample. The dearth of star formation before $\sim 60$ Myr ago could partly reflect the loss of dissolved low-mass clusters, although such a sharp drop is not naturally explained by the gradual disruption expected for typical bound open clusters \citep{almeida_open_2024} and may therefore be at least partly real. Over the last 40 Myr, the present-day 1 kpc cube sample has an average SFR of $823~\mathrm{M}_{\odot}~\mathrm{Myr}^{-1}$. Restricting the same history to the traced 1 kpc cube gives an average \(\Sigma_{\mathrm{SFR}}\) of $894~\mathrm{M}_{\odot}~\mathrm{Myr}^{-1}~\mathrm{kpc}^{-2}$. The apparent drop over the last $\sim 5$ Myr should also be interpreted cautiously, since many very young clusters may still be embedded and therefore underrepresented in Gaia-selected cluster catalogs \citep{hunt_selection_2026}.

The ccSN rate rises gradually after the SFR increases, as expected from the delay between cluster formation and the first explosions in the stellar-lifetime model. Over the last 40 Myr, the corresponding average ccSN rate is $7.7~\mathrm{Myr}^{-1}$ for the present-day 1 kpc cube sample, while the average \(\Sigma_{\mathrm{ccSNe}}\) is $7.4~\mathrm{Myr}^{-1}~\mathrm{kpc}^{-2}$ in the traced box. The latter corresponds to a mean interval of $\Delta t_{\rm ccSN}\simeq0.14$ Myr between ccSNe within the traced area. The first row of Table~\ref{tab:rates_clusters_ob} reports these present-day 1 kpc cube averages over the $1~\mathrm{kpc}^{2}$ comparison area and gives a Milky Way-equivalent rate of $0.55\pm0.03$ century$^{-1}$. The ccSN curves are smoother than the SFR curves because explosions at a given time can be supplied by clusters formed across a range of earlier epochs, with the progenitor lifetimes spreading each burst of star formation into a broader feedback episode. The shaded bands in the ccSN panels show the 16th--84th percentile range across IMF realizations.

The strongest recent SFR peaks in the annotated panel are not produced by single clusters. They are produced by multiple regional or family contributors, including Orion, Sco--Cen as the nearby young component of the $\alpha$~Per family, the $\gamma$~Vel family, Cr135, and M6. These formation episodes produce broader ccSN rate enhancements over the following tens of Myr.

\subsection{Present-day OB star comparison}\label{sec:results_ob_compare_draft}

Table~\ref{tab:rates_clusters_ob} also reports the rates inferred from the Q25 and ALS III OB-star subsamples. The Q25 subsample contains 164 stars with Q25-inferred masses of at least $8\,\mathrm{M}_{\odot}$ within the same volume as our 1 kpc cube cluster sample. Because stars above this mass have lifetimes of up to approximately 40 Myr, these present-day counts primarily trace star formation and ccSN production over the last $\sim40$ Myr. We therefore compare them with the cluster rates averaged over the same interval. Q25 gives $\Sigma_{\mathrm{SFR}}\simeq900\,\mathrm{M}_{\odot}\,\mathrm{Myr}^{-1}\,\mathrm{kpc}^{-2}$, $\Sigma_{\mathrm{ccSNe}}=10\,\mathrm{Myr}^{-1}\,\mathrm{kpc}^{-2}$, and a Milky Way-equivalent rate of $0.71$ century$^{-1}$, factors of 1.1 and 1.3 above the cluster-based local rates. The Q25 rate also closely matches the value reported by \citet{quintana_census_2025}, while our cluster-based rate lies within the range inferred from the Hunt \& Reffert cluster catalog by \citet{quintana_how_2025}. The ALS III subsample contains 648 stars assigned $\texttt{Cat}=\texttt{M}$ within the same volume, giving $\Sigma_{\mathrm{SFR}}\simeq3700\,\mathrm{M}_{\odot}\,\mathrm{Myr}^{-1}\,\mathrm{kpc}^{-2}$, $\Sigma_{\mathrm{ccSNe}}=39\,\mathrm{Myr}^{-1}\,\mathrm{kpc}^{-2}$, and a Milky Way-equivalent rate of $2.78$ century$^{-1}$, factors of 4.4 and 5.0 above the cluster-based local rates. Therefore, Q25 implies rates similar to those inferred from the clusters, whereas ALS III implies substantially higher rates.

\begin{figure}[t]
    \centering
    \includegraphics[width=0.90\columnwidth]{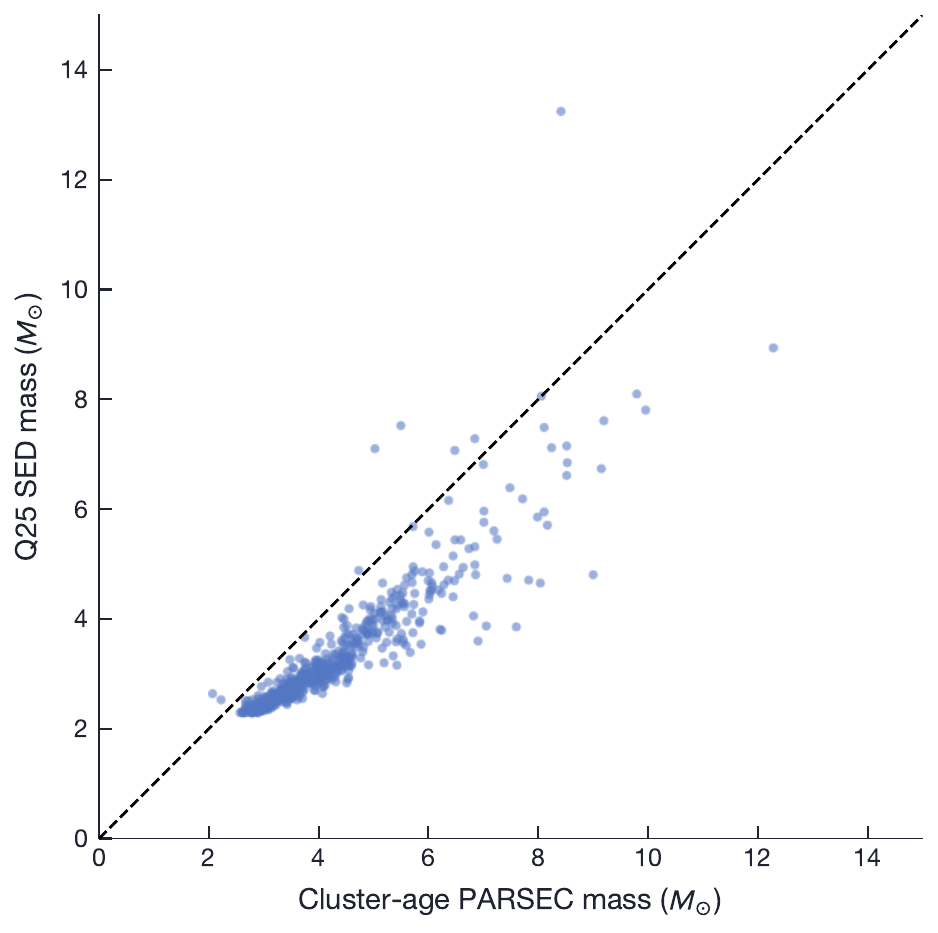}
    \caption{Q25 SED mass versus cluster-age PARSEC mass for 635 stars in 197 clusters within the 1 kpc cube, including Q25 sources above and below $8\,M_{\odot}$. The dashed line marks equality; the median cluster-to-Q25 mass ratio is 1.3.}
    \label{fig:q25_cluster_isochrone_mass_comparison}
\end{figure}

As described in the Methods, we crossmatch our cluster members with all available Q25 and ALS III stars in the same volume, regardless of whether they belong to our primary OB-star subsamples. This gives cluster-based masses for 635 Q25 stars and 97 ALS III stars. By including stars both inside and outside the Q25 $M\geq8\,M_{\odot}$ and ALS III $\texttt{Cat}=\texttt{M}$ selections, we can check for lower-mass contaminants and missed massive stars. The resulting effective counts and rates are reported in the ``OB stars (Isochrone mass comparison)'' rows of Table~\ref{tab:rates_clusters_ob}. Figure~\ref{fig:q25_cluster_isochrone_mass_comparison} compares the Q25 masses with our isochrone-based masses for the 635 matched stars. Our masses are larger by a median factor of 1.3. Applying this factor to the Q25 masses increases the number of stars above $8\,M_{\odot}$ from 164 to 219. This gives $\Sigma_{\mathrm{SFR}}\simeq1200\,M_{\odot}\,\mathrm{Myr}^{-1}\,\mathrm{kpc}^{-2}$, $\Sigma_{\mathrm{ccSNe}}=13\,\mathrm{Myr}^{-1}\,\mathrm{kpc}^{-2}$, and a Milky Way-equivalent ccSN rate of $0.94$ century$^{-1}$. For ALS III, 58 of the 73 matched $\texttt{Cat}=\texttt{M}$ stars have isochrone masses below $8\,M_{\odot}$, giving a contamination fraction of 79\%. None of the 24 matched non-$\texttt{M}$ stars exceed this mass, so we identify no missed massive stars among the small set of matched sources outside the selection. If the measured contamination and selection incompleteness fractions apply to the full ALS III subsample, its effective count decreases from 648 to 133 stars. This gives $\Sigma_{\mathrm{SFR}}\simeq800\,M_{\odot}\,\mathrm{Myr}^{-1}\,\mathrm{kpc}^{-2}$, $\Sigma_{\mathrm{ccSNe}}=8\,\mathrm{Myr}^{-1}\,\mathrm{kpc}^{-2}$, and a Milky Way-equivalent ccSN rate of $0.57$ century$^{-1}$. The isochrone comparison raises the Q25 rates but lowers the ALS III rates dramatically, bringing the two catalog estimates closer to one another. Only the corrected ALS III estimate approaches the cluster-based value.

The spectroscopic classifications provide a second, more direct check on the two photometric selections. The resulting rates are reported in the ``OB stars (Spectral class comparison)'' rows of Table~\ref{tab:rates_clusters_ob}. Among the spectroscopically classified Q25 stars, 25 of the 127 stars above the Q25 mass cut are non-OB stars, giving a contamination fraction of 20\%. Q25 also excludes 125 of the 227 spectroscopically identified massive stars, giving a selection incompleteness fraction of 55\%. Combining these fractions gives a count correction factor of 1.8 and increases the effective Q25 count from 164 to 293 stars. The corresponding rates are $\Sigma_{\mathrm{SFR}}\simeq1700\,M_{\odot}\,\mathrm{Myr}^{-1}\,\mathrm{kpc}^{-2}$, $\Sigma_{\mathrm{ccSNe}}=18\,\mathrm{Myr}^{-1}\,\mathrm{kpc}^{-2}$, and $R_{\rm MW}=1.26$ century$^{-1}$. For ALS III, 110 of the 234 spectroscopically classified $\texttt{Cat}=\texttt{M}$ stars are non-OB stars, giving a contamination fraction of 47\%. ALS III also excludes 158 of the 282 spectroscopically identified massive stars, giving a selection incompleteness fraction of 56\%. Combining these fractions gives a count correction factor of 1.2 and increases the effective ALS III count from 648 to 781 stars. The corresponding rates are $\Sigma_{\mathrm{SFR}}\simeq4400\,M_{\odot}\,\mathrm{Myr}^{-1}\,\mathrm{kpc}^{-2}$, $\Sigma_{\mathrm{ccSNe}}=47\,\mathrm{Myr}^{-1}\,\mathrm{kpc}^{-2}$, and $R_{\rm MW}=3.35$ century$^{-1}$. For ALS III, the spectral-class check gives an effective count and ccSN rate almost six times larger than the isochrone check (781 versus 133 stars; $47$ versus $8\,\mathrm{Myr}^{-1}\,\mathrm{kpc}^{-2}$), showing that the inferred correction depends strongly on the adopted diagnostic.

The disagreement between Q25 and ALS III is also apparent at the source level. After applying the same volume cuts, the two subsamples share only 60 stars out of 731 unique sources. Appendix Figure~\ref{fig:massive_star_catalog_comparison} shows that the Q25, ALS III, and reconstructed cluster populations also have different projected spatial distributions.

The isochrone and spectral class comparisons give substantially different rate estimates, particularly for ALS III, preventing us from confidently determining how incomplete the cluster-based reconstruction is. We therefore do not use either comparison to renormalize the reconstructed ccSN map and discuss the implications for the absolute local star formation and ccSN rates further in \S\ref{sec:discussion_rates}.

\subsection{Time dependence of ccSN clustering}\label{sec:results_clustering_draft}

\begin{figure*}[!tp]
    \centering
    \includegraphics[width=0.82\textwidth]{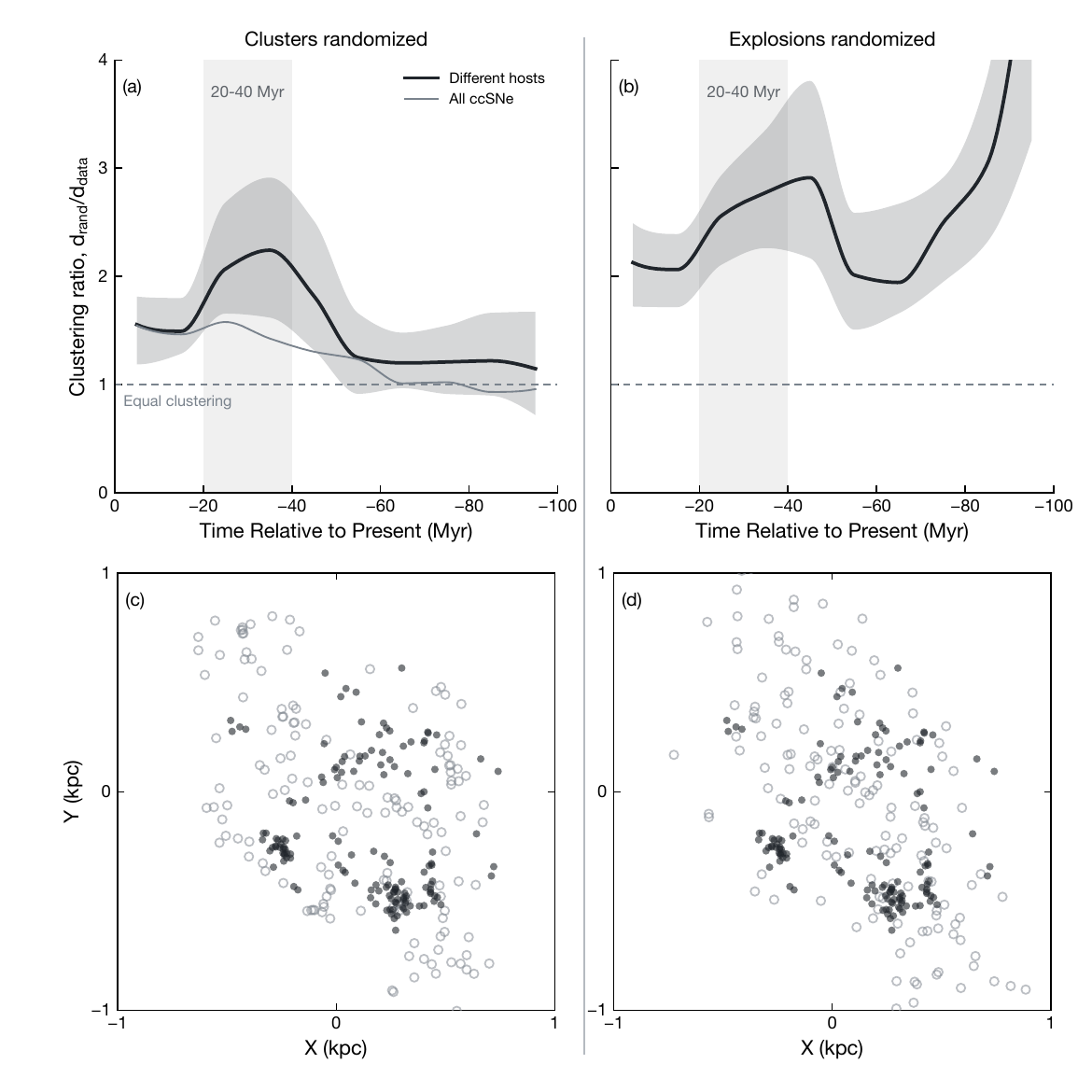}
    \caption{Projected ccSN clustering over the last 100 Myr for the 312 clusters in the present-day 1 kpc cube sample. Panels (a) and (b) show the randomized-cluster and randomized-location comparisons, respectively. Both report the nearest-neighbor ratio $d_{\rm rand}/d_{\rm data}$, where values above unity indicate stronger clustering in the reconstruction. The main curve in panel (a) includes only pairs from different hosts; the thin solid line includes all ccSN pairs and is shown without uncertainty shading. Panel (b) includes all pairs. Smooth lines interpolate the median ratios between independent 10 Myr bins, and shaded bands interpolate the corresponding 16th--84th percentiles across paired IMF and randomized-map realizations. Panels (c) and (d) show representative paired maps for 20--40 Myr ago. Filled dark points mark reconstructed ccSNe; open gray circles mark the corresponding randomized comparison. All catalogs are matched in event count and evaluated over the same traced area.}
    \label{fig:clustering_evolution}
\end{figure*}

Figure~\ref{fig:clustering_evolution} compares the reconstructed ccSNe with two randomized samples. In both panels, we report the ratio of the median nearest-neighbor distances, $d_{\rm rand}/d_{\rm data}$, in independent 10 Myr bins. Values above unity mean that the reconstructed ccSNe are more spatially clustered than the randomized sample.

Panel (a) tests whether the reconstructed cluster hosts are more concentrated than expected from a randomized cluster catalog. We randomize the present-day cluster positions and velocities while retaining the cluster properties used to generate their ccSNe. Multiple explosions therefore remain associated with the same cluster in both the reconstructed and randomized samples. The thick curve considers only pairs of supernovae from different clusters and measures the clustering among distinct hosts. The thin curve also includes pairs of supernovae from the same cluster.

The reconstructed cluster hosts are more concentrated throughout the last 50 Myr. The different-host ratio is approximately 1.5--1.6 over the last 20 Myr and 1.8--2.2 from 20 to 50 Myr ago. When pairs from the same cluster are included, the ratio is approximately 1.5 over the last 20 Myr and 1.3--1.6 from 20 to 50 Myr ago. Because repeated explosions from the same cluster remain grouped in both samples, including these pairs reduces the contrast between the reconstructed and randomized cluster distributions.

Panel (b) tests the total clustering of the ccSNe. We retain the reconstructed explosion times but redistribute the individual events among randomized cluster trajectories. This breaks both the grouping of explosions from the same cluster and the spatial correlations among different clusters. The resulting ratio is approximately 2.1 over the last 20 Myr and 2.6--2.9 from 20 to 50 Myr ago. Together, the two comparisons show that the reconstructed clustering reflects both the compact past arrangement of distinct cluster hosts and repeated explosions within individual clusters. Because the null models randomize different parts of the reconstruction, the difference between their ratios should not be interpreted as a quantitative separation of these two contributions.

\section{Discussion}\label{sec:discussion}

    Our reconstructed ccSN history offers a way to connect recent massive star feedback to the present-day 3D structure of the nearby ISM. Recent work has used 3D dust maps to resolve kiloparsec-scale structures, including the superclouds identified by \citet{kormann_superclouds_2025}, and combined them with H~I and CO emission to estimate line-of-sight gas motions \citep{soler_kinetic_2025,mccallum_milky_2026}. While we do not carry out a detailed comparison between our reconstructed ccSNe and the nearby gas structures and motions, we discuss their relation to known shells and bubbles, their spatial clustering, and uncertainties in the absolute local star formation and ccSN rates.

\subsection{Mapping feedback input to nearby ISM shells and bubbles}\label{sec:discussion_ism}

    The strongest recent ccSN enhancements in Figure~\ref{fig:recent_map} coincide with or lie near present-day dust cavities, including the Local Bubble, Orion--Eridanus, IVS/Gum, GSH~238, and Cepheus Far. While the ccSN map averages activity over the last 15 Myr and the dust map shows the present-day ISM, their spatial correspondence identifies the reconstructed feedback sites most relevant to the nearby shell and bubble network.

    The Local Bubble is the low-density cavity surrounding the Sun, first identified through soft X-ray emission and interstellar absorption and later mapped in 3D gas and dust \citep{cox_local_1987,snowden_rosat_1997,lallement_3d_2003}. Its origin has long been linked to feedback from Sco--Cen: \textit{Hipparcos}-era traceback studies estimated that roughly 10--20 SNe were required to form the Bubble and associated this feedback with nearby Sco--Cen subgroups \citep{maiz-apellaniz_origin_2001,berghofer_origin_2002,fuchs_search_2006}. More recently, \citet{zucker_star_2022} used \textit{Gaia}-era stellar populations and 3D dust maps to infer that the Local Bubble began expanding about 14 Myr ago and triggered star formation on its surface. \citet{oneill_local_2024} mapped the Bubble as an irregular 3D dust shell with a northern extension associated with the Local Chimney, which we show projected in Figure \ref{fig:recent_map}. We reconstruct \(9^{+2}_{-3}\) ccSNe in the Local Bubble volume over the last 15 Myr (Table~\ref{tab:shell_counts}), slightly lower but consistent with these earlier estimates and with the \(11^{+3}_{-3}\) SNe from young $\alpha$~Per family members found by \citet{swiggum_most_2024}. 

    We reconstruct \(17^{+5}_{-4}\) ccSNe within Orion--Eridanus over the last 15 Myr (Table~\ref{tab:shell_counts}), making it a prominent recent enhancement in Figure~\ref{fig:recent_map}. Including stellar winds, our reconstructed sources provide a mechanical-energy input of \(1.9^{+0.5}_{-0.5}\times10^{52}\) erg within the same aperture, before accounting for energy losses or coupling to the surrounding gas. This is similar to the \(1.8^{+1.5}_{-0.4}\times10^{52}\) erg estimated for Orion OB1, which is sufficient to explain the size and expansion of the H~I shell \citep{brown_orion_1995,voss_probing_2010}.

    The nested shells in Orion--Eridanus point to several episodes of feedback \citep{ochsendorf_nested_2015}. \textit{Gaia}-era studies likewise identify distinct stellar populations ranging in age from a few Myr to about 20 Myr \citep{kounkel_apogee-2_2018,zari_structure_2019,kos_discovery_2019,chen_discovery_2020}. Radial cluster expansion among the roughly 6--10 Myr populations and the expanding dust cavity associated with Barnard's Loop provide further evidence for this extended history \citep{swiggum_evidence_2021,foley_3d_2023}. Although \citet{kos_galah_2021} found no detectable chemical enrichment from supernovae, our reconstruction predicts \(9^{+3}_{-3}\) ccSNe within the Orion--Eridanus aperture 8--21 Myr ago. These explosions may therefore have left no measurable abundance signature in the sampled young stars.

    Vela is the most prominent recent enhancement in Figure~\ref{fig:recent_map}, with \(41^{+7}_{-7}\) reconstructed ccSNe enclosed by the \(17~{\rm Myr}^{-1}~{\rm kpc}^{-2}\) contour over the last 15 Myr (Table~\ref{tab:contour_counts}). This region is also the most structurally complex part of the nearby map, because several shells, stellar populations, and cluster families overlap along similar sightlines. The largest associated structure is GSH 238+00+09, marked in Figure~\ref{fig:recent_map}. GSH 238 is a kiloparsec-scale supershell, visible in modern 3D dust maps as one of the largest nearby cavities, with an inferred age of order 25--30 Myr \citep{heiles_whence_1998,swiggum_most_2024}. Within the broader GSH 238 environment lies the IRAS Vela Shell (IVS), which \citet{gao_origin_2025} mapped in 3D using the \citet{edenhofer_parsec-scale_2024} dust map. They found a shell diameter of 138 pc and a vertical extent of roughly 140 pc, and showed that the IVS is spatially correlated with the Gum Nebula. Previous work has also associated the Gum Nebula, the IRAS Vela Shell, and the surrounding H~I/radio-continuum structures with recent massive star feedback \citep{woermann_kinematics_2001,testori_radio_2006}.

    Our reconstructed ccSNe history suggests that the Vela region has been shaped by feedback over an extended time interval. In GSH 238, we reconstruct \(132^{+11}_{-11}\) ccSNe over the last 30 Myr (Table~\ref{tab:shell_counts}). This large number is dominated by older activity from the Cr135 and M6 cluster families. \citet{swiggum_most_2024} linked GSH 238 primarily to the Cr135 family, whose member clusters trace back to a compact volume consistent with a past massive star-forming complex and are now dispersed throughout the present-day shell. Figure~\ref{fig:timeline} shows that the M6 family also contributes substantially to the same region, with the M6 and Cr135 ccSN enhancements beginning to overlap by 10--20 Myr ago.

    The IRAS Vela Shell (IVS) is a dust shell associated with the ionized Gum Nebula in Vela, the most prominent recent ccSN enhancement in our map. Within the 3D IVS volume, we reconstruct \(2^{+2}_{-1}\) ccSNe over the last 3 Myr from clusters in the Cr135 and $\gamma$~Vel families. This agrees with the one to two SNe required to supply the shell's momentum, while the ring of young $\gamma$~Vel/Vela OB2 stars around the IVS suggests a shared history \citep{cantat-gaudin_ring_2019,gao_origin_2025}. Notably, the broader region is also host to the well-known Vela supernova remnant, further evidence of very recent supernova activity in Vela, although the remnant lies in the foreground of the IVS \citep{cha_distance_1999}.

    Several additional recent ccSN enhancements in Figure~\ref{fig:recent_map} are labeled in the contour map. Cepheus Far overlaps a region with extensive shell and star formation literature, including the nearby Cepheus Flare shell and the more distant Cepheus Bubble around Cep OB2, both of which have been discussed in connection with expanding gas, triggered star formation, and massive star feedback \citep{olano_interstellar_2006,kiss_star_2006,kun_pre-main-sequence_2009,abraham_morphology_2000,szilagyi_gaia_2021,szilagyi_gaia_2023}. The Ara/Norma/Circinus enhancement overlaps the Sagittarius Spur Extension identified by \citet{kormann_superclouds_2025}, which extends the Sagittarius Spur first mapped by \citet{kuhn_high_2021} into the fourth Galactic quadrant toward Ara OB1 and Circinus \citep{pantaleoni_gonzalez_alma_2025}. The same sightlines contain several distance and velocity components, including the foreground H~I shell GSH~337+00$-$05 and more distant molecular material associated with Ara OB1 \citep{mcclure-griffiths_galactic_2002,arnal_12co_2003,henderson_interstellar_2008}. The Aquila Rift enhancement overlaps the Serpens/Aquila star-forming complex, where W40 and Serpens South show localized expanding molecular shells and dense filamentary star formation \citep{shimoikura_cluster_2019,shimoikura_detailed_2020,maury_formation_2011,konyves_census_2015}. The Lacerta enhancement lies near Lac OB1 and associated remnant clouds such as LBN~437 and GAL~110$-$13, which have been interpreted as feedback-shaped sites of possible triggered star formation \citep{olano_molecular_1994,lee_triggered_2007,chen_lacerta_2008,kaltcheva_lacerta_2009}.


\subsection{Clustered ccSNe}\label{sec:discussion_clustered}

    We found that the reconstructed ccSNe are more spatially clustered than expected for a random distribution of ccSNe (Figure \ref{fig:clustering_evolution}). This clustering arises partly because massive clusters can host multiple ccSNe, and partly because related clusters trace back to more compact configurations with their siblings in the past, especially within the Cr135, M6, and $\alpha$~Per families (Figure \ref{fig:timeline}).
    
    The importance of clustered massive star feedback for the ISM has been recognized for decades: early bubble models showed that continuous supernovae injection can inflate expanding cavities, while H~I shell catalogs demonstrated that large shells and supershells structure the Galactic ISM on scales of hundreds of parsecs \citep{castor_interstellar_1975,weaver_interstellar_1977,heiles_hi_1979}. Subsequent work connected these structures directly to OB associations and clustered supernovae, showing that sequential explosions can drive superbubbles, chimneys, and disk breakout \citep{tomisaka_sequential_1981,maclow_superbubbles_1988}.
    
    Recent simulations of clustered stellar feedback and superbubble evolution have shown that the spatial and temporal coherence of supernovae strongly affects how feedback couples to the ISM. Clustered explosions can maintain hot cavities, drive expanding shells, and promote breakout or outflows, although the retained energy and momentum depend sensitively on ambient density, explosion cadence, mixing, and whether the explosions remain confined within the disk \citep{kim_superbubbles_2017,gentry_enhanced_2017,gentry_momentum_2020,fielding_clustered_2018,el-badry_evolution_2019,orr_bursting_2022}. 
    
    The Local Bubble discussed in \S\ref{sec:discussion_ism} provides nearby evidence that clustered supernova feedback can produce vertical breakout. \citet{oneill_local_2024} find that its 3D dust shell is not closed and has a chimney-like morphology towards the north Galactic pole. Similar breakout signatures may accompany other ccSN enhancements in our map. Future work combining the reconstructed event histories with 3D dust and gas morphology could test which regions remained confined as shells and which opened into chimneys or larger-scale outflows.

\subsection{A Bursty, Recent Local Star Formation History}
\label{sec:bursty}
 
The reconstructed star formation history in the top-left panel of Figure~\ref{fig:local_history} has implications for a long-standing question: whether star formation proceeds as a quasi-steady, secular process or through discrete bursts. This distinction has implications for the efficiency of stellar feedback and the lifecycle of the local ISM. A secular history would imply that recent star formation is regulated by large-scale Galactic dynamics and the mean gas surface density \citep{kennicuttevans2012,krumholztan2007}, yielding a smoothly varying rate. In contrast, a bursty history points to rapid, feedback-regulated cycling between molecular clouds, star formation, and feedback \citep{kruijssen_fast_2019}.
 
Our reconstruction resolves the recent local star formation history finely enough to distinguish between these modes and favors the bursty interpretation. Averaged over the past 40~Myr, the present-day 1 kpc cube sample forms stars at $823~M_\odot~\mathrm{Myr}^{-1}$. The strongest bursts reach $\Sigma_{\mathrm{SFR}}\simeq2.5\times10^{3}~M_\odot~\mathrm{Myr}^{-1}~\mathrm{kpc}^{-2}$, roughly three times the long-term average, and rise and fall over only a few~Myr, comparable to the 5--7~Myr molecular-cloud lifetimes measured in Milky Way-mass simulations \citep{benincasa2020}. These peaks are not artifacts of individual clusters: the dominant episodes are built from multiple regional or family contributors, including Orion, the young Sco--Cen component of the $\alpha$~Per family, the $\gamma$~Vel family, Cr135, and M6, with the average rate beginning to climb $\sim$45~Myr ago as these families formed. The amplitude of the peaks relative to the low mean therefore favors a bursty rather than secular star formation history in the Solar Neighborhood.
 
We caution that the burst amplitudes are subject to the limitations discussed above. The small volume makes the rate sensitive to stochastic sampling of a handful of massive complexes. Incomplete sampling of independent star-forming regions can increase the apparent variation in the star formation rate on small spatial scales \citep{kruijssenlongmore2014}. Very young ($\lesssim$$5~\mathrm{Myr}$) clusters may still be embedded and underrepresented, while the apparent dearth before $\sim$60~Myr is at least partly a selection effect (\S\ref{sec:results_rates}). Distributed star formation outside the surviving cluster population could also raise the mean relative to the peaks. None of these effects, however, erases the resolved peak-to-trough structure, which is driven by the timing of distinct, independently dated cluster families.
 
This bursty, feedback-regulated picture is reinforced by two independent signatures in our reconstruction. The strongest bursts correspond closely to the present-day shells and cavities carved into the local ISM (\S\ref{sec:discussion_ism}). In addition, \citet{fauchergiguere2018} predict that bursty star formation should be accompanied by particularly strong spatio-temporal clustering of supernovae. This agrees with our reconstruction: the ccSNe remain significantly more clustered than the randomized comparisons over the reconstructed interval (\S\ref{sec:discussion_clustered}). We therefore interpret the top-left panel of Figure~\ref{fig:local_history} as spatially resolved evidence that, over the past $\sim$50~Myr, recent cluster-traced star formation in the Solar Neighborhood proceeded through discrete episodes rather than a quasi-steady buildup.

\subsection{The absolute star formation and ccSN rates in the Solar Neighborhood}\label{sec:discussion_rates}

    As discussed in \S\ref{sec:results_ob_compare_draft}, the rates reconstructed from the 1 kpc cube sample would be lower limits if a substantial fraction of recent star formation occurred outside the identified clusters. We test this by comparing our results with two independent \textit{Gaia} DR3-based OB star catalogs. Stars with \(M>8\,M_\odot\) live for \(\lesssim40\) Myr and trace both clustered and dispersed populations, providing a separate estimate of the recent star formation and ccSN rates. However, the two catalogs overlap only weakly despite targeting nearly the same massive star population, so they cannot provide a unique normalization. This disagreement leaves uncertain how much star formation occurs in compact clusters, how quickly clusters disperse after gas removal, and how much massive star feedback the nearby ISM has received.
    
    The Q25-based star formation surface density is only slightly higher than the cluster-based value (Table~\ref{tab:rates_clusters_ob}). This agrees with \citet[hereafter QHP25]{quintana_how_2025}, who found that compact clusters younger than 10 Myr account for most, but not all, of the local star formation rate inferred from Q25 OB stars. For comparison, \citet{lada_embedded_2003} estimated \(\Sigma_{\mathrm{SFR}}=1000\text{--}3000~M_\odot~{\rm Myr}^{-1}~{\rm kpc}^{-2}\) from embedded clusters in the Solar Neighborhood. The QHP25, Q25-based, and cluster-based estimates all lie near the lower end of this range.

    QHP25 and the Q25 sample used in our work probe different timescales. QHP25 use the full catalog, including late-B stars that can live for several hundred Myr, whereas we include only stars with \(M>8\,M_\odot\). These stars trace the past \(\sim40\) Myr and are direct ccSN progenitors. QHP25 conclude that most stars form in compact or bound clusters. This differs from the hierarchical picture, in which bound clusters are the densest parts of a continuous star-forming distribution and contain only a minority of newly formed stars \citep{elmegreen_variations_2008,kruijssen_fraction_2012,ward_not_2018}. Over the timescale relevant to our reconstruction, the Q25 sample used in our work suggests that the 1 kpc cube sample misses only a modest fraction of recent star formation. Our reconstructed ccSN counts would therefore be only slightly underestimated.

    ALS III suggests a different picture, unless its massive-star sample is highly contaminated. Its star formation surface density is 4.4 times the cluster-based value and lies just above the upper end of the embedded-cluster range from \citet{lada_embedded_2003} (Table~\ref{tab:rates_clusters_ob}). Lada \& Lada (2003) attributed the difference between embedded- and open-cluster rates to the rapid dispersal of young clusters after feedback removes their natal gas. If the same process explains the ALS III excess, some clusters must lose their compact structure within \(\lesssim10\) Myr. Their massive stars would remain in ALS III after the clusters were no longer identifiable. The high ALS III rate could also support a more hierarchical picture in which much of the recent star formation lies outside recognizable compact or bound clusters.

    The ALS III-based estimate of \(\Sigma_{\mathrm{ccSNe}}\) is 5.0 times higher than our cluster-based value (Table~\ref{tab:rates_clusters_ob}). Previous studies of nearby massive stars have also found relatively high rates. \citet{tammann_galactic_1994} inferred \(29~{\rm Myr}^{-1}~{\rm kpc}^{-2}\), close to the ALS III value, while \citet{reed_new_2005} found a somewhat lower massive-star birthrate. \citet{grenier_gamma-ray_2000} estimated \(75\text{--}95~{\rm Myr}^{-1}~{\rm kpc}^{-2}\). The ALS III-based rate is therefore high, but falls within the range of previous local estimates.

    Our cluster reconstruction gives a Milky Way-equivalent ccSN rate of \(0.55\pm0.03~\mathrm{century}^{-1}\). This is close to the \(0.4\text{--}0.5~\mathrm{century}^{-1}\) inferred by \citet{quintana_census_2025}, but lower than estimates of \(1.63\pm0.46~\mathrm{century}^{-1}\) and \(1.9\pm1.1~\mathrm{century}^{-1}\) from other Galactic measurements \citep{rozwadowska_rate_2021,diehl_radioactive_2006}. Our Q25 and ALS III comparisons give \(0.71\) and \(2.78~\mathrm{century}^{-1}\), respectively (Table~\ref{tab:rates_clusters_ob}). The cluster and Q25 rates lie near the lower end of previous estimates, while the ALS III rate lies near the upper end.

    Regional comparisons in Sco--Cen and Orion point toward the higher rates suggested by ALS III. In Sco--Cen, SigMA recovers \(98.3\%\) of the HR23 members but identifies about 3.5 times as many young stars overall \citep{ratzenbock_significance_2023,ratzenbock_star_2023}. Preliminary results for Orion likewise recover \(91.7\%\) of the corresponding Hunt \& Reffert members but identify about 3.4 times as many stars (A. Rottensteiner et al., in preparation). Both comparisons suggest that compact clusters contain only part of the young stellar population, supporting the higher star formation and ccSN rates inferred from ALS III.

    On the other hand, the present-day ISM cautions against adopting the ALS III catalog normalization. \citet{swiggum_most_2024} found that the supernova counts inferred for their HR23-based cluster families are already broadly consistent with the sizes of major 3D dust cavities, including the Local Bubble and GSH~238+00+09. \citet{soler_kinetic_2025} recently combined the 3D dust map of \citet{edenhofer_parsec-scale_2024} with H~I and CO line emission, reconstructed the local line-of-sight gas velocity field, and estimated the kinetic-energy excess in nearby ISM structures. Adopting a 10\% coupling efficiency between supernova blast waves and the ISM, this excess requires roughly 120 supernovae, fewer than but roughly consistent with the \(>200\) inferred from the cluster families of \citet{swiggum_most_2024}. \citet{mccallum_three-dimensional_2025} provide a separate constraint from the ionized gas, finding that higher-SFR simulations over-disrupt and over-ionize the local ISM. Thus, while ALS III may indicate missing distributed massive stars, adopting the ALS III-implied rates would likely overestimate the feedback required by the present-day local ISM.

    Taken together, these comparisons point to a broader open problem: the absolute rate of recent local star formation remains uncertain, and different tracers give different answers. This is important since the adopted rate sets the amount of massive star feedback available to shape the present-day ISM. They suggest:
    \begin{itemize}
        \item Our cluster-based ccSN rates should be interpreted as conservative lower limits, since some recent massive star formation may occur outside the identified cluster population.

        \item The missing component is not uniquely constrained: some methods imply only modest incompleteness, while others allow for a much larger dispersed population.

        \item A substantially higher normalization remains possible in a hierarchical picture where many young stars form in low-density associations or rapidly disperse from compact clusters.

        \item However, the present-day ISM does not clearly require such a high normalization and may already be broadly consistent with the cluster-based feedback budget.

    \end{itemize}

\section{Summary and Conclusions}\label{sec:conclusions}
    We have reconstructed the recent ($<50$ Myr), core-collapse supernova history in the solar neighborhood by tracing backwards the Galactic orbits of 568 young star clusters. We re-derived their ages and masses, then propagated uncertainties in age, mass, and orbital trajectories, along with stochastic IMF sampling, to construct a probabilistic, time-resolved map of where ccSNe likely occurred throughout the solar neighborhood.

    The reconstructed ccSN history over the past 15 Myr reveals strong enhancements in Orion, Vela, and Cepheus-Far. Many enhancements are located inside known cavities and shells traced by the 3D dust distribution -- particularly inside the Orion-Eridanus shell, the Local Bubble, GSH 238+00+09, and the IRAS-Vela Shell. Our estimated ccSN counts within these cavities are consistent with previous estimates of the energy needed to have powered their formation. 

    The local star formation and ccSN rates were dominated by the cluster families of \citet{swiggum_most_2024} over the past 50 Myr, which emerge initially in our reconstruction as compact ccSN enhancements that spread over time. Notably, the ccSN enhancements of the Cr135 and M6 families merged approximately 20 Myr ago, and the smaller $\gamma$~Vel family (Vela OB2) began to form and produce ccSNe shortly afterward in the same Vela region.

    We find a bursty star formation history in the Solar Neighborhood, with sharp peaks reaching \(\Sigma_{\mathrm{SFR}}\simeq2.5\times10^3~M_\odot~{\rm Myr}^{-1}~{\rm kpc}^{-2}\), roughly three times the 40 Myr average of \(894~M_\odot~{\rm Myr}^{-1}~{\rm kpc}^{-2}\). The average rate began rising around 45 Myr ago due to the onset of the $\alpha$~Per and M6 cluster families. The ccSN rate rose shortly after and sustained \(\Sigma_{\mathrm{ccSNe}}\simeq8.7~{\rm Myr}^{-1}~{\rm kpc}^{-2}\) from 5 to 25 Myr ago.

    We also find that local ccSNe were spatially clustered, rather than randomly distributed, over the past 50 Myr. This clustering partly reflects the fact that individual clusters can host multiple ccSNe close together in space and time. The stronger clustering 20–40 Myr ago, however, also reflects the initially compact configuration of sibling clusters forming within the three cluster families.

    The absolute recent local star formation and ccSN rates remain uncertain, with possible normalization factors ranging from modest, \(\sim 1.5\times\), to large, \(\sim 5\times\), depending on the adopted OB star catalog. We therefore treat our reconstructed rates as lower limits and use this range to bracket the plausible recent feedback history of the Solar Neighborhood. Future spectroscopic targeting of candidate massive OB stars will be essential for narrowing this uncertainty.

    Our reconstructed ccSN history is intended as a data-driven framework for connecting recent massive star feedback to the present-day structure of the nearby ISM. It may also serve as a useful input for numerical simulations attempting to model solar neighborhood-like ISM conditions \citep{walch_silcc_2015,kim_superbubbles_2017,rathjen_silcc_2021,kim_introducing_2023,mccallum_persistence_2024}. Additionally, studying the solar system's past trajectory through our ccSN reconstruction might help connect to terrestrial, lunar, and cosmic-ray records of supernova-produced radionuclides such as \(^{60}\mathrm{Fe}\) \citep{wallner_recent_2016,breitschwerdt_locations_2016,fields_near-earth_2019,schulreich_numerical_2017,schulreich_numerical_2023,miller_heliospheric_2022,maconi_solar_2025}. Looking ahead, the upcoming fourth \textit{Gaia} Data Release will provide improved parallaxes and proper motions, together with many additional radial velocities, enabling richer young cluster catalogs whose orbits can be traced backward more precisely. Complementary radial velocity coverage from SDSS-V and other upcoming spectroscopic surveys will also improve the 3D kinematics of young stars.

\section*{Data Availability}
The data files supporting this work will be available in an updated version of this manuscript.

\begin{acknowledgments}
    C.Z. and C.S. acknowledge support from NSF CAREER Award No. 2442546 (CAREER: Charting the Formation, Transformation, and Destinies of Gas and Young Stars in the Solar Neighborhood).
This work has made use of data from the European Space Agency (ESA) mission \textit{Gaia} (\url{https://www.cosmos.esa.int/gaia}), processed by the \textit{Gaia} Data Processing and Analysis Consortium (DPAC, \url{https://www.cosmos.esa.int/web/gaia/dpac/consortium}). Funding for the DPAC has been provided by national institutions, in particular the institutions participating in the \textit{Gaia} Multilateral Agreement.
GPT-5.6 was used to assist with manuscript grammar and organization, for selected analysis tasks, and to develop the interactive figure in its entirety. The authors reviewed and verified all generated text, code, and analysis outputs and take full responsibility for the manuscript.
\end{acknowledgments}

\software{Astropy \citep{astropy_collaboration_astropy_2022}, dustmaps \citep{green_dustmaps_2018}, emcee \citep{foreman_mackey_emcee_2013}, galpy \citep{bovy_galpy_2015}, imf (\url{https://github.com/keflavich/imf}), Matplotlib \citep{hunter_matplotlib_2007}, NumPy \citep{harris_array_2020}, pandas \citep{mckinney_data_2010}, and SciPy \citep{virtanen_scipy_2020}.}


\newpage
\restartappendixnumbering
\appendix

\section{Supplementary comparisons}\label{sec:appendix_comparisons}

\subsection{Cluster age and mass comparison}\label{sec:appendix_age_mass_comparison}

\begin{figure*}[t]
    \centering
    \includegraphics[width=\textwidth]{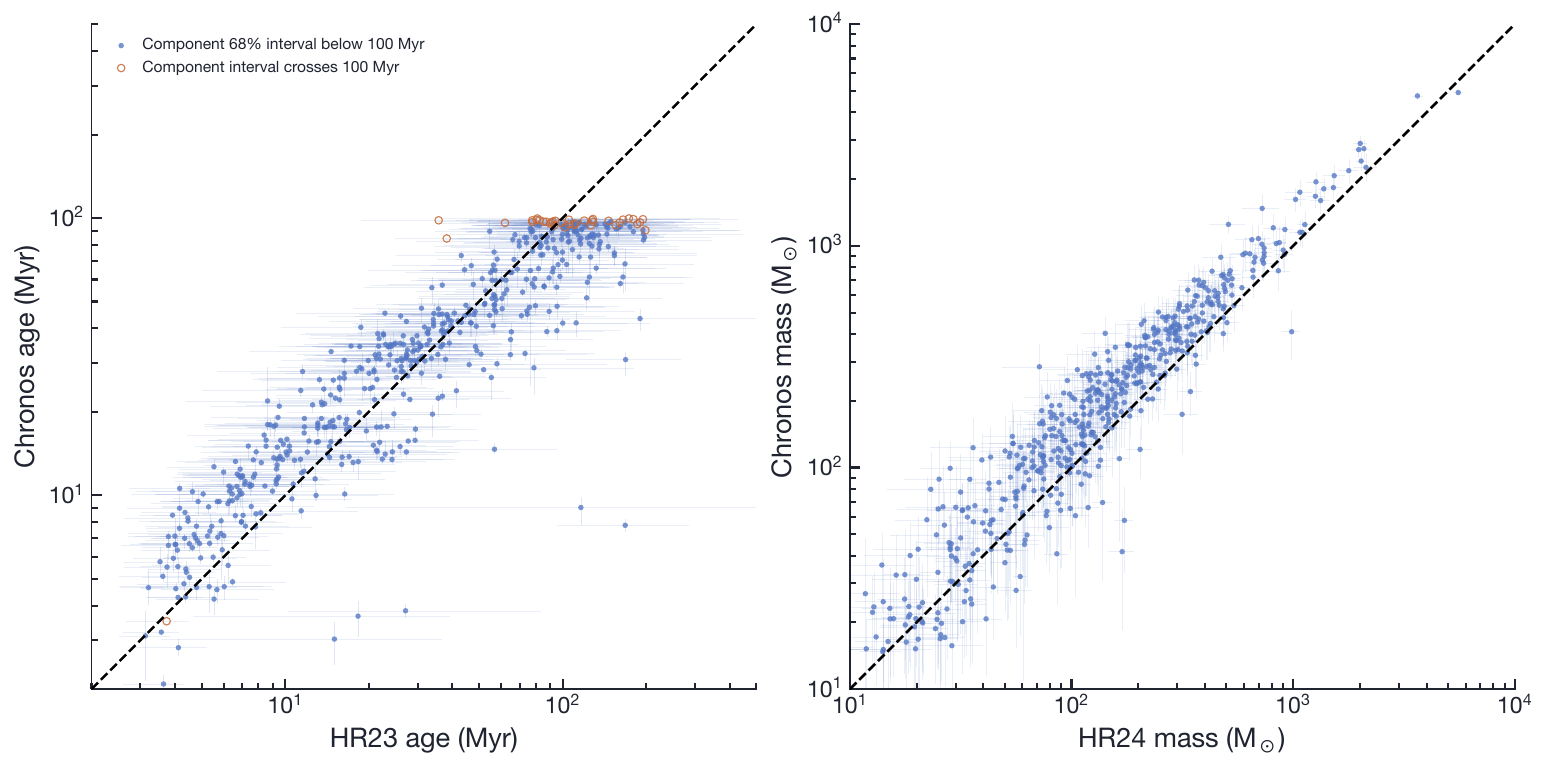}
    \caption{Comparison between the HR23 catalog values and the re-derived \texttt{Chronos}/PARSEC values for clusters in the 2.5 kpc cube sample. The left panel compares cluster ages, with HR23 ages on the horizontal axis and the median of the selected \texttt{Chronos} posterior component on the vertical axis. Vertical error bars show the conditional 68\% interval of that component. Filled points have intervals that remain below 100 Myr; open points have selected-component intervals that cross the 100 Myr sample boundary. The right panel compares HR23 \texttt{mass\_all} values to the inferred initial cluster masses after applying the Almeida et al. (2024) correction.}
    \label{fig:age_mass_comparison}
\end{figure*}

Figure~\ref{fig:age_mass_comparison} compares the HR23 catalog parameters with the re-derived \texttt{Chronos}/PARSEC values adopted in this work. This comparison is intended as a diagnostic of the inputs to the ccSN reconstruction: HR23 catalog ages define the initial young cluster candidate sample, while the re-derived ages and masses define the 2.5 kpc cube sample, star formation history, and IMF sampling.

\subsection{Spatial comparison of local massive star catalogs}\label{sec:appendix_massive_stars}

\begin{figure*}[t]
    \centering
    \includegraphics[width=\textwidth]{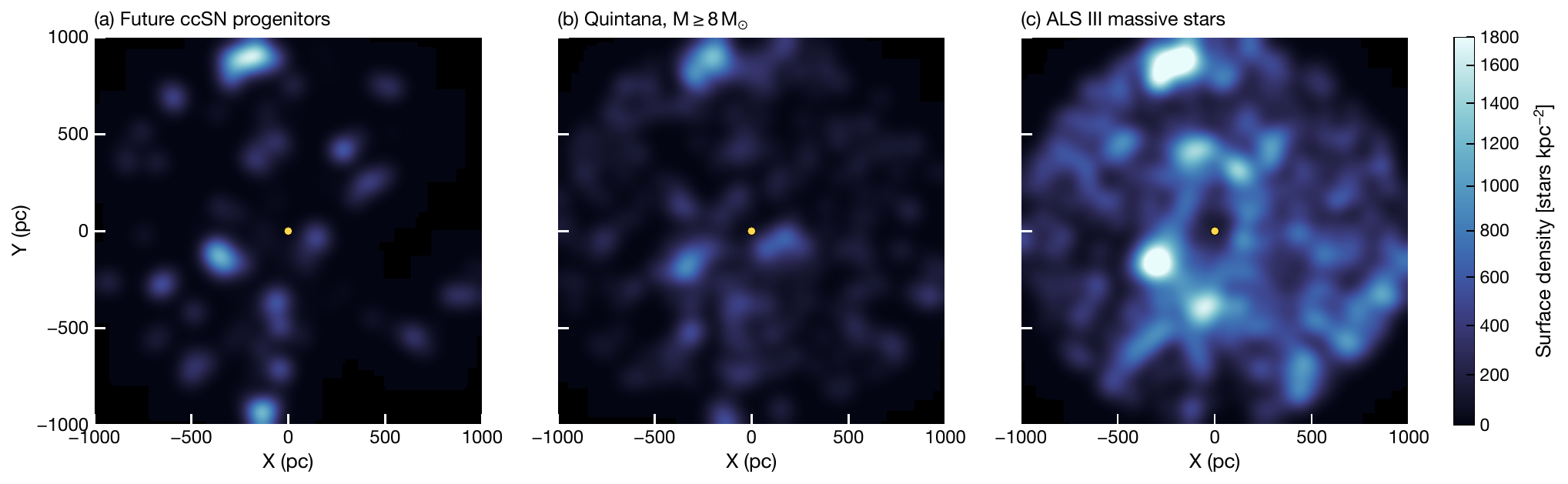}
    \caption{Comparison of the present-day projected distribution of nearby massive stars within 1 kpc of the Sun and $|Z| < 500$ pc. Panel (a) shows the median inferred surface-density distribution of stars in the 1 kpc cube sample that are expected to explode as core-collapse supernovae in the future, plotted at their present-day positions. Panel (b) shows the Quintana catalog restricted to stars with $M \geq 8~M_{\odot}$. Panel (c) shows the corresponding distribution of stars flagged as massive (\texttt{Cat = M}) in the ALS III catalog using \texttt{Dist\_P50} distances. All three panels are displayed with the same KDE-style rendering and spatial footprint for direct comparison.}
    \label{fig:massive_star_catalog_comparison}
\end{figure*}

Figure~\ref{fig:massive_star_catalog_comparison} compares the projected Galactic \(X\)--\(Y\) distributions used for the massive star normalization check. The comparison uses the wider view described in \S\ref{sec:data_ob_catalogs}, within 1 kpc of the Sun and \(|Z|<500\) pc, so that differences in the spatial distributions of the reconstructed surviving massive stars, ALS III massive star candidates, and Q25 massive star candidates can be seen directly.

\subsection{2.5 kpc cube sample}\label{sec:appendix_xy1250_history}

\begin{figure*}[t]
    \centering
    \includegraphics[width=\textwidth]{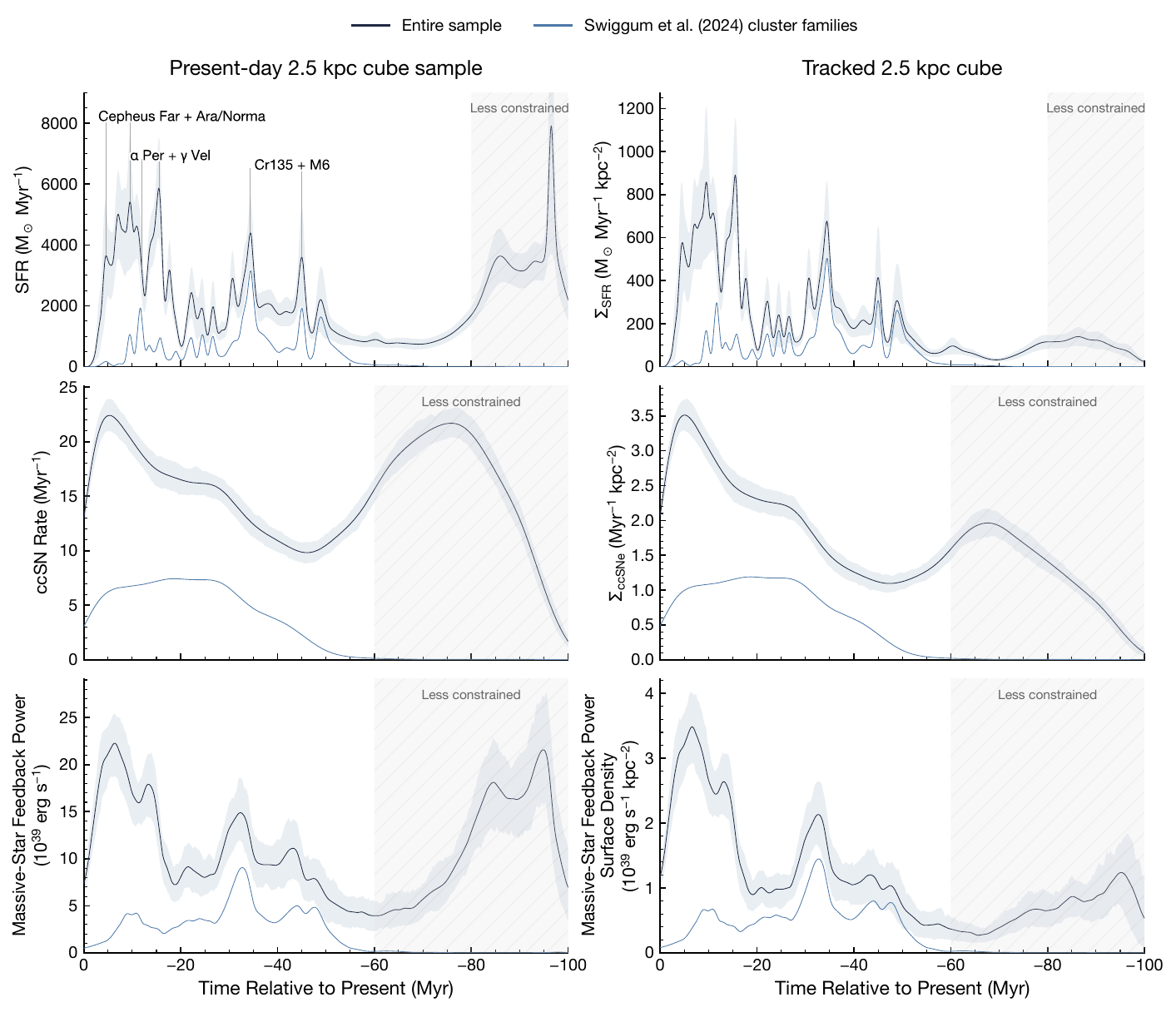}
    \caption{Same as Figure~\ref{fig:local_history}, but for the 2.5 kpc cube sample. Labels in the upper-left panel identify selected cluster family and regional contributions to the recent star formation peaks. We do not label peaks in the less-constrained intervals.}
    \label{fig:xy1250_history}
\end{figure*}

Figure~\ref{fig:xy1250_history} shows the star formation, ccSN, and massive-star feedback power histories for the 2.5 kpc cube sample. The left column includes all activity associated with clusters whose present-day positions satisfy $|X|,|Y|,|Z|\leq1.25$ kpc. The right column includes only activity inside the corresponding co-moving, co-rotating cube at each lookback time and reports surface densities over its $6.25~\mathrm{kpc}^{2}$ projected area.

\bibliography{references}
\bibliographystyle{aasjournalv7}

\end{document}